\PassOptionsToPackage{unicode}{hyperref}
\PassOptionsToPackage{hyphens}{url}
\PassOptionsToPackage{dvipsnames,svgnames,x11names}{xcolor}
\documentclass[
  letterpaper,
  DIV=11,
  numbers=noendperiod]{scrartcl}

\usepackage{amsmath,amssymb}
\usepackage{lmodern}
\usepackage{iftex}
\ifPDFTeX
  \usepackage[T1]{fontenc}
  \usepackage[utf8]{inputenc}
  \usepackage{textcomp} % provide euro and other symbols
\else % if luatex or xetex
  \usepackage{unicode-math}
  \defaultfontfeatures{Scale=MatchLowercase}
  \defaultfontfeatures[\rmfamily]{Ligatures=TeX,Scale=1}
\fi
\IfFileExists{upquote.sty}{\usepackage{upquote}}{}
\IfFileExists{microtype.sty}{% use microtype if available
  \usepackage[]{microtype}
  \UseMicrotypeSet[protrusion]{basicmath} % disable protrusion for tt fonts
}{}
\makeatletter
\@ifundefined{KOMAClassName}{% if non-KOMA class
  \IfFileExists{parskip.sty}{%
    \usepackage{parskip}
  }{% else
    \setlength{\parindent}{0pt}
    \setlength{\parskip}{6pt plus 2pt minus 1pt}}
}{% if KOMA class
  \KOMAoptions{parskip=half}}
\makeatother
\usepackage{xcolor}
\ifx\paragraph\undefined\else
  \let\oldparagraph\paragraph
  \renewcommand{\paragraph}[1]{\oldparagraph{#1}\mbox{}}
\fi
\ifx\subparagraph\undefined\else
  \let\oldsubparagraph\subparagraph
  \renewcommand{\subparagraph}[1]{\oldsubparagraph{#1}\mbox{}}
\fi

\providecommand{\tightlist}{%
  \setlength{\itemsep}{0pt}\setlength{\parskip}{0pt}}\usepackage{longtable,booktabs,array}
\usepackage{calc} % for calculating minipage widths
\usepackage{etoolbox}
\makeatletter
\patchcmd\longtable{\par}{\if@noskipsec\mbox{}\fi\par}{}{}
\makeatother
\IfFileExists{footnotehyper.sty}{\usepackage{footnotehyper}}{\usepackage{footnote}}
\makesavenoteenv{longtable}
\usepackage{graphicx}
\makeatletter
\def\maxwidth{\ifdim\Gin@nat@width>\linewidth\linewidth\else\Gin@nat@width\fi}
\def\maxheight{\ifdim\Gin@nat@height>\textheight\textheight\else\Gin@nat@height\fi}
\makeatother
\setkeys{Gin}{width=\maxwidth,height=\maxheight,keepaspectratio}
\makeatletter
\def\fps@figure{htbp}
\makeatother
\newlength{\cslhangindent}
\newlength{\csllabelwidth}
\newlength{\cslentryspacingunit} % times entry-spacing
\newenvironment{CSLReferences}[2] % #1 hanging-ident, #2 entry spacing
 {% don't indent paragraphs
  \setlength{\parindent}{0pt}
  \ifodd #1
  \let\oldpar\par
  \def\par{\hangindent=\cslhangindent\oldpar}
  \fi
  \setlength{\parskip}{#2\cslentryspacingunit}
 }%
 {}
\usepackage{calc}

\usepackage{booktabs}
\usepackage{caption}
\usepackage{longtable}
\usepackage{colortbl}
\usepackage{array}
\KOMAoption{captions}{tableheading}
\makeatletter
\@ifpackageloaded{caption}{}{\usepackage{caption}}
\AtBeginDocument{%
\ifdefined\contentsname
  \renewcommand*\contentsname{Table of contents}
\else
  \newcommand\contentsname{Table of contents}
\fi
\ifdefined\listfigurename
  \renewcommand*\listfigurename{List of Figures}
\else
  \newcommand\listfigurename{List of Figures}
\fi
\ifdefined\listtablename
  \renewcommand*\listtablename{List of Tables}
\else
  \newcommand\listtablename{List of Tables}
\fi
\ifdefined\figurename
  \renewcommand*\figurename{Figure}
\else
  \newcommand\figurename{Figure}
\fi
\ifdefined\tablename
  \renewcommand*\tablename{Table}
\else
  \newcommand\tablename{Table}
\fi
}
\@ifpackageloaded{float}{}{\usepackage{float}}
\floatstyle{ruled}
\@ifundefined{c@chapter}{\newfloat{codelisting}{h}{lop}}{\newfloat{codelisting}{h}{lop}[chapter]}
\floatname{codelisting}{Listing}

\makeatother
\makeatletter
\@ifpackageloaded{caption}{}{\usepackage{caption}}
\@ifpackageloaded{subcaption}{}{\usepackage{subcaption}}
\makeatother
\makeatletter
\@ifpackageloaded{tcolorbox}{}{\usepackage[many]{tcolorbox}}
\makeatother
\makeatletter
\@ifundefined{shadecolor}{\definecolor{shadecolor}{rgb}{.97, .97, .97}}
\makeatother
\ifLuaTeX
  \usepackage{selnolig}  % disable illegal ligatures
\fi
\IfFileExists{bookmark.sty}{\usepackage{bookmark}}{\usepackage{hyperref}}
\IfFileExists{xurl.sty}{\usepackage{xurl}}{} % add URL line breaks if available
\hypersetup{
  pdftitle={The SIDO Performance Model for League of Legends},
  pdfauthor={Amy X. Zhang; Parth Naidu},
  colorlinks=true,
  linkcolor={blue},
  filecolor={Maroon},
  citecolor={Blue},
  urlcolor={Blue},
  pdfcreator={LaTeX via pandoc}}

\title{The SIDO Performance Model for League of Legends}
\author{Amy X. Zhang \and Parth Naidu}
\date{2/2/24}

\begin{document}
\maketitle
\begin{abstract}
League of Legends (LoL) has been a dominant esport for a decade, yet the
inherent complexity of the game has stymied the creation of analytical
measures of player skill and performance. Current industry standards are
limited to easy-to-procure individual player statistics that are
incomplete and lacking context as they do not take into account teamplay
or game state. We present a unified performance model for League of
Legends which blends together measures of a player's contribution within
the context of their team, insights from traditional sports metrics such
as the Plus-Minus model, and the intricacies of LoL as a complex team
invasion sport. Using hierarchical Bayesian models, we outline the use
of gold and damage dealt as a measure of skill, detailing players'
impact on their own-, their allies'- and their enemies' statistics
throughout the course of the game. Our results showcase the model's
increased efficacy in separating professional players when compared to a
Plus-Minus model and to current esports industry standards, while metric
quality is rigorously assessed for discrimination, independence, and
stability. Readers might also find additional qualitative analytics
which explore champion proficiency and the impact of collaborative
team-play. Future work is proposed to refine and expand the SIDO
performance model, offering a comprehensive framework for esports
analytics in team performance management, scouting and research realms.
\end{abstract}
\ifdefined\Shaded\renewenvironment{Shaded}{\begin{tcolorbox}[breakable, interior hidden, enhanced, frame hidden, borderline west={3pt}{0pt}{shadecolor}, boxrule=0pt, sharp corners]}{\end{tcolorbox}}\fi

\renewcommand*\contentsname{Table of contents}
{
\hypersetup{linkcolor=}
\setcounter{tocdepth}{3}
\tableofcontents
}
\hypertarget{introduction}{%
\section{Introduction}\label{introduction}}

League of Legends, a seminal multiplayer online battle arena (MOBA)
game, has taken the world by storm, driving the rapid and explosive
growth of professional esports. Over the past decade, competitive gaming
has evolved from a niche hobby to a global phenomenon with stadiums
filled to capacity, millions of viewers tuning in online, and players
competing at the highest level for substantial prizes. As the esports
ecosystem continues to expand, the pursuit of excellence in competitive
gaming becomes increasingly paramount.

Despite the meteoric rise of competitive gaming, the longevity and
multigenerational status of League of Legends and other similar titles
is still up for debate and hinges on our ability to develop efficient
learning, training, and performance methods to sustainably engage every
part of the ecosystem. Achieving this goal requires a twofold approach:
the practical implementation of programs catering to players of all
skill levels and a continuous commitment to pioneering research.
Innovation in both practical training programs and academic research is
essential for advancing the esports industry and cultivating talent that
can compete at the highest echelons of competitive gaming.

Yet, a significant roadblock stands in the way of progress---the field
of esports analytics remains in its infancy, particularly concerning the
critical domain of player performance analysis. While traditional sports
have benefited from decades of statistical modeling and performance
metrics, the world of esports is only beginning to scratch the surface
of what can be achieved through data-driven insights. The lack of a
standardized and quantifiable model for player performance hampers not
only research efforts but also the exploration of alternative training
methodologies.

A robust quantitative model of player performance is required to build a
more informed and innovative future for competitive gaming. This paper
presents first steps towards bridging the gap by holistically
considering a player's total contribution to their team, not just their
individual performance. This is similar in concept to a Plus-Minus model
from traditional sports, and we compare our method to both a Plus-Minus
model and to a simple average model. We hope to provide a foundation for
the evolution of esports analytics and the optimization of player
performance in League of Legends and beyond.

\hypertarget{background}{%
\section{Background}\label{background}}

\hypertarget{performance-metrics-in-traditional-sports}{%
\subsection{Performance metrics in traditional
sports}\label{performance-metrics-in-traditional-sports}}

The application of statistical analysis to sports has transformed how
teams evaluate player performance and make strategic decisions. One
prominent example that has left an indelible mark on the world of sports
analytics is ``Moneyball'' (Lewis 2004). This famous approach,
popularized by the book and subsequent film, showcases how statistical
analysis can revolutionize the understanding of player contributions in
sports.

Moneyball's origins are in baseball, an atomistic sport that offers a
rich array of individual statistics for precise player assessment, such
as On-Base Percentage (OBP). OBP measures a player's ability to reach
base safely, including through hits, walks, and hit-by-pitches, relative
to their total plate appearances. Precise metrics like OBP allow teams
to quantify and compare player contributions accurately, fostering
strategic decisions in player recruitment and game tactics.

In contrast, complex team invasion sports, like football or basketball,
present several challenges (Gerrard 2007):

\begin{itemize}
\item
  Complexity: These sports involve a wide range of player actions,
  making it challenging to isolate individual contributions.
\item
  Interdependence: Player actions are highly interdependent, with
  cascading effects on their teammates, making it hard to attribute
  success solely to individuals.
\item
  Joint Actions: Players often collaborate on actions like tackling,
  further complicating individual attribution.
\item
  Continuity: Some sports have continuous play (e.g., soccer), while
  others are segmented (e.g., American football), impacting the analysis
  of player actions.
\item
  Specialization: Players' roles vary widely, from all-around players in
  soccer to highly specialized units in American football.
\end{itemize}

To perform detailed player performance analysis in these complex
invasion sports, two key measurement problems must be addressed (Gerrard
2007):

\begin{itemize}
\item
  Attribution Problem: Correctly allocating individual contributions to
  joint and interdependent actions. Deciding how much credit each player
  receives for shared actions is a challenge.
\item
  Weighting Problem: Determining the significance of each player action
  in influencing the overall match outcome. Different approaches, either
  subjective based on expert judgment or statistical based on regression
  analysis, can be used to assign weights.
\end{itemize}

However, the hierarchical structure of complex invasion sports
introduces complexity in assigning weights, as higher-level actions
(e.g., scoring) depend on lower-level actions (e.g., ball possession and
passing). This hierarchical structure requires expert judgment to
determine relative weightings between different levels.

\hypertarget{league-of-legends-lol-as-a-complex-team-invasion-sport}{%
\subsection{League of Legends (LoL) as a complex team invasion
sport}\label{league-of-legends-lol-as-a-complex-team-invasion-sport}}

In League of Legends (LoL), gameplay centers around strategic team-based
competition within a virtual battlefield called the Summoner's Rift. Two
teams, each consisting of five players, engage in intense battles with
the primary objective of destroying the enemy team's Nexus, located
within their base, while defending their own. This marks LoL as a team
invasion sport, with traditional sports analogues such as basketball and
football.

Players control one of 160+ unique characters known as \emph{champions},
each possessing different abilities and belonging to distinct types such
as tanks, damage dealers, and support characters. Players also select
which role in the game they play, designated as \emph{top},
\emph{middle}, \emph{bot}, \emph{support}, or \emph{jungle}. These roles
correspond both to a high-level playstyle (for example, bot players
typically play fragile damage dealers) and to a zone of the battlefield
the player is responsible for. The combination of different player roles
and champion types allows for diverse team compositions and strategies.

Players enhance their champions' power by purchasing \emph{items} using
the gold they earn. Effective communication, teamwork, and strategic
decision-making are essential for success. Champions' abilities must be
utilized wisely and players must work together to outmaneuver opponents
and gain control of the battlefield.

While there are no set rules for gameplay at any time point of the game,
the game is colloquially often split into the early game, mid-game, and
late-game. Each of these game phases is typified by different styles of
play, based on the state of the battlefield, or game map.

In the early game, all players start with equal gold and resources and,
for the most part, adhere to a standardized set of actions. This is due
to the state of the map in the early game--each team must defend the
boundaries of their territory by protecting defensive \emph{towers}
placed along lanes on the map. This results in the top, middle, bot, and
support players facing off against their lane opponents in the middle of
the map, between two sets of opposing towers. The only significant
variation at this stage comes from the jungle role, the only role which
does not go to a tower at the beginning of the game and instead roams in
the jungle area between lanes. The jungler's decisions about pathing and
map impact introduce some unpredictability. During this period, the
focus is primarily on individual performance, with players striving to
gain personal advantages over their lane opponents. This is also often
called the \emph{laning phase} of the game.

The middle phase of the game introduces more complexity and a wider
range of decisions due to three key developments:

\begin{enumerate}
\def\labelenumi{\arabic{enumi}.}
\item
  Tower dynamics: Protective barriers (\emph{tower plates}) around
  towers are reduced, which opens up more strategic options for players
  in those lanes. Towers may also have been destroyed, which alters the
  initial lane setups.
\item
  Player strength disparity: As the game progresses, disparities in
  player strength emerge across different roles, influenced by how
  players have fared against their counterparts.
\item
  Neutral objectives: This period also sees the appearance of the
  \emph{herald}, a neutral objective that can provide a significant
  advantage. Securing the herald aids in obtaining tower plates and
  enhances map control.
\end{enumerate}

While laning (i.e., directly facing your lane opponent in an isolated
position on the map like in the early game) is often present during the
mid game, this phase serves as a transition to broader map strategies,
with players balancing the acquisition of resources for themselves and
their team. Skirmishes become more common, typically involving smaller
groups and occurring more sporadically.

By late game, the laning phase has concluded and all tower plates have
been removed. The choices available to players now are influenced by the
outcomes of the previous phases. This stage is characterized by more
extensive \emph{teamfights} focused around major objectives. Teamfights
tend to involve all players, who typically operate in closer proximity
to one another, making strategic decisions with the entire team's
interests in mind.

LoL is a prime example of a complex team invasion sport, with inherent
complexities that extend beyond those found in typical invasion sports:

\begin{enumerate}
\def\labelenumi{\arabic{enumi}.}
\item
  Diversity of champions and items: LoL boasts a staggering roster of
  over 160 champions, each with unique abilities and roles.
  Additionally, the game features more than 200 items that players can
  use to enhance their champions' abilities or adapt to the evolving
  match circumstances. This diversity results in an astronomical number
  of potential team compositions and strategic choices. In essence, the
  myriad of champions and items adds an extra layer of complexity,
  making it challenging to analyze and attribute individual player
  contributions accurately.
\item
  Continuity and game state dependency: Unlike traditional sports, where
  each game state starts from a relatively clean slate, LoL's continuity
  means that every game state is intricately linked to the one preceding
  it. Each player accrues gold and resources at different rates based on
  their decisions and the events of previous moments in the game. This
  interdependence between game states and the cumulative nature of
  resources and advantages earned over time introduces significant
  complexities in assessing individual player contributions. Actions in
  the early stages can have a profound impact on the late-game, further
  complicating the analysis.
\item
  Frequent game changes: Riot Games, the developer of LoL, introduces
  regular updates and changes to the game, including champions, items,
  and game mechanics. These adjustments occur every few weeks, creating
  a dynamic and evolving gaming environment. While these updates keep
  the game fresh and exciting, they pose a significant challenge for
  creating and training statistical models. The frequent changes make it
  difficult to establish stable and consistent datasets over extended
  periods. This, in turn, hampers the ability to analyze trends, making
  long-term statistical analysis and prediction more complex.
\end{enumerate}

A glossary of terms is provided in the Appendix,
Section~\ref{sec-glossary}.

\hypertarget{industry-standards}{%
\section{Industry standards}\label{industry-standards}}

A number of issues currently exist within the industry for evaluating
player contributions. This is in part because of the complexity of data
collection and analysis, as well as the intricate nature of the
game---the vast pool of champions to select from, frequent game changes,
and continuity of game states. The issues are:

\begin{itemize}
\item
  Simplified metrics: Data aggregation platforms such as Oracle's Elixir
  and Gol.gg provide average statistics like damage dealt, gold earned
  per minute, and creep score (CS) differentials. These metrics are
  often discussed and used to gauge player performance. However, these
  numbers are typically presented in isolation, without comprehensive
  consideration of various contextual factors such as game length, team
  compositions, match-ups, and significant in-game events.
\item
  Selective use of metrics: Due to the recognition that these simple
  metrics do not fully encapsulate a player's performance, they are only
  used selectively. Analysts and enthusiasts in the League of Legends
  community recognize the limitations of these statistics and the need
  to contextualize them within the broader scope of the game
  (MissFortuneDaBes 2022; Hajir 2021; Baker 2018; LoL Esports 2017).
  This selective approach to metric usage reflects a consensus that
  individual performance in League of Legends cannot be distilled down
  to a few basic numbers.
\item
  Perception of player performance: Given the high attribution problem
  in League of Legends, separating individual performances from match
  outcomes can be challenging. As a result, the perception of a player's
  performance is often influenced by the team's overall success.
  Analysts and viewers are more inclined to view players from winning
  teams as performing better. This trend is particularly evident during
  post-season awards, where there is a notable correlation between the
  team with the most regular-season wins and the number of
  representatives from that team who receive recognition as top
  performers in their respective roles. In the the summer of 2023,
  Korean teams KT and GenG had the first and second most regular season
  wins, respectively, and each of their players were voted first and
  second in their role\footnote{LCK Spring All-Pro awards:
    https://www.oneesports.gg/league-of-legends/t1-players-all-lck-first-team/.
    LCK Summer All-Pro awards:
    https://www.oneesports.gg/league-of-legends/all-lck-first-team-summer-2023/.}.
  Far from an anomaly, this is part of a long-standing trend in
  post-season awards.
\end{itemize}

In essence, the industry's approach to evaluating player performance in
League of Legends is a pragmatic one. Recognizing the complexities
involved in attributing individual contributions to match outcomes, the
community has adopted simplified metrics as a starting point. However,
these metrics are regarded as insufficient on their own, leading to
their selective use alongside qualitative analysis and expert opinions.
Ultimately, the perception of player performance is often influenced by
the team's success, reflecting the challenging nature of separating
individual excellence from the dynamics of a team-based esport like
League of Legends.

Individual statistics, while prominent and readily available, only
represent one dimension of a player's multifaceted skill expression. In
the highly dynamic and strategic environment of team invasion games,
players frequently find themselves at crossroads where choices that
enhance the team's chances of victory may not align with optimizing
their individual metrics. These crucial decisions often go beyond the
realm of industry-standard statistics.

Consider a scenario where a player willingly sacrifices their lane
advantage to roam and support a struggling teammate in a different lane.
While this selfless act may not bolster their individual gold or damage
statistics, it significantly contributes to their team's overall
success. Similarly, a player who strategically engages an enemy,
disrupting their actions and abilities in a team skirmish, may not
emerge with impressive damage numbers, yet their impact on the game, by
inhibiting the enemy's effectiveness, is immeasurable.

The focus on individual performance has given birth to many tools,
particularly in solo queue, a game mode where individual players compete
in matches without pre-arranged teams. A core part of solo queue is the
ranked ladder, which consists of various tiers and divisions reflecting
a player's competitive standing relative to all other players on the
server. Websites like OP.GG provide tools to assess player performance
in solo queue. These platforms calculate a performance score, typically
on a scale from 0 to 10, by analyzing a player's various metrics such as
vision score, KDA ratio, damage dealt, and more and comparing them to
their teammates. More dedicated players seeking to use these tools to
improve may then focus on their individual performance over contributing
to the team.

In order to comprehensively capture a player's performance, we have to
value decisions they make in the game that they believe to be in service
of winning the game, notably how effectively the player helps his
teammates or inhibits the enemy players. For example, if we're
considering gold as a primary resource, a player's performance should
consider not only how efficient they are at collecting gold, but how
their actions help their teammates gather gold or prevent the enemy in
comparison to others in the same position.

\hypertarget{related-work}{%
\section{Related work}\label{related-work}}

In solo queue, current research on player skill has, for the most part,
defined player skill using Elo (Elo 1967) or Elo-based models. These
models define more skilled players as players who win more games.
Sapienza, Goyal, and Ferrara (2019) used the TrueSkill ranking system
(Herbrich, Minka, and Graepel 2007) to measure how teammates influence a
player's skill improvement in Defense of the Ancients 2 (DotA). Aung et
al. (2018) studied the relationship between early learning rate and long
term performance by predicting end-of-season rank based on the rate of
change in MMR at the beginning of the season. Dehpanah et al. (2021)
compared different methods of aggregating win-based skill metrics to
derive a team skill metric for matchmaking. Other papers focus on
evaluation and improvement of motor skills (Toth et al. 2021;
Velichkovsky et al. 2019; Jeong et al. 2022).

Many times, the success of a performance metric is measured by whether
the metric can predict wins. Xia, Wang, and Zhou (2019) use kills as a
performance metric in DotA, which they found to be more related to
strategic/tactical awareness than mechanics. Bahrololloomi et al. (2022)
use xp per minute.

This focus on wins is partially rooted in the ethos of
competition---what matters most, after all, is whether the team wins. A
rather more prolific body of literature in recent years exists around
predicting wins (Birant and Birant 2022; Liang 2021; Do et al. 2021;
D.-H. Kim, Lee, and Chung 2020; X. Zhang et al. 2017). However, this
line of thinking fails to separate the result of the game from
performance. A player can perform well but still lose; likewise, another
player can perform badly but still win (Kahn and Williams 2016).

Applying this thinking to competitive gaming quickly leads to logical
inconsistencies when evaluating individual player performance. In
traditional sports, player-level performance is separated from
team-level performance through methods such as win probability added
(WPA) and plus/minus models (Hvattum 2019; Pettigrew 2015; Kaplan,
Mongeon, and Ryan 2014). Within esports, consideration of how to
separate player-level performance from team performance is still in its
nascent stages. Clark, Macdonald, and Kloo (2020) demonstrated the
Plus-Minus model's utility in DotA. Suznjevic, Matijasevic, and Konfic
(2015) argued for publishers to include consideration of in-game
performance metrics when designing matchmaking systems, rather than only
team result.

MOBA games, and competitive gaming in general, includes many players who
take more defensive or supportive roles. In these cases, their
performance is best measured in terms of impact on their team or on the
enemy team. Few works consider player skill within a team context.
Dehpanah et al. (2021) treat team skill as an aggregate measure of
individual skill. Team skill on its own remains difficult to quantify,
though attempts have been made. Y. J. Kim et al. (2017) use collective
intelligence, a group measure of intelligence, to predict performance
among ranked teams in League of Legends. Birant and Birant (2022) used
post-game statistics and multi-instance learning to predict team wins.
Kahn and Williams (2016) showed that transactive memory systems strongly
predict the likelihood of a team winning a game in League of Legends.
Drachen et al. (2014) used intra-team spatial distance to measure skill
level in DotA, ranging from novice to pro.

In traditional sports, performance metrics such as the Plus-Minus
statistic seek to measure a player's skill level in terms of their
impact on their team. The method we present has the same conceptual base
as the Plus-Minus statistic, where a player's skill level is measured
within the context of the impact on their team. Our method is also able
to take into account other external variables such as player role and
champion, both key factors which drastically impact how a player is
evaluated, and we demonstrate the advantages of our method through
quantitative comparison with the Plus-Minus model of Clark, Macdonald,
and Kloo (2020).

\hypertarget{data}{%
\section{Data}\label{data}}

Solo queue data were collected from the Riot API\footnote{Riot API:
  https://developer.riotgames.com/docs/lol} on a daily basis from
accounts ranked Diamond and above.

Solo queue data represent a valuable resource outside of competitive
game data for two reasons: one, larger sample size, as the number of
games a pro player plays in solo queue is typically much greater than
the number of competitive matches they play; two, semi-random game
matching allows for more easily attributing player actions to player
skill. For example, in a competitive game, the bot player may have the
highest gold in the game at 15 minutes because their jungler or mid
laner are consistently visiting their lane to help them kill enemies and
gain gold. How much of the bot player's gold can be attributed to their
own play, versus attributed to the mid laner or jungler's actions?
Competitive teams often form typical patterns of play such as this,
which makes attribution difficult.

In solo queue, players are randomly matched to other players with
roughly the same Elo. Most games have a unique set of players and
champions in the game. The semi-random shuffling of ally and enemy
players means the confounding issue that exists in competitive play is
less of an issue in solo queue.

\hypertarget{criteria-for-game-inclusion}{%
\subsection{Criteria for game
inclusion}\label{criteria-for-game-inclusion}}

We considered ranked solo queue games played by high-ranking accounts in
the North America (NA), Korea (KR), and EU-West (EUW) servers for games
played on patches 13.14 - 13.18. These patches roughly correspond to
games played from July 18, 2023 to September 27, 2023, and mark the
beginning of a major patch release by Riot.

Accounts on each server are separated into tiers based on their rank,
updated daily. High-ranking accounts were identified as the top 1000
accounts on each server based on \emph{LP} at the end of each day, which
includes the vast majority of professional players' accounts. This
corresponds to the \emph{grandmaster} (players ranked top 1000-301) and
\emph{challenger} (players ranked in the top 300) tiers. Any given
player may have multiple high-ranking accounts within a server.

Faulty internet or a lack of desire to play the game may cause a player
to `disconnect', taking them out of the game. As these games were found
to skew model results, we removed them by excluding games where any
single player took less than 500 damage by the 7-minute mark.

Each of the five roles in the game (top, jungle, mid, bot, and support)
requires a different set of skills and so it is common for players to
``main'' one or two roles and only occasionally play roles outside of
their main roles. Within each role, we identified a set of accounts who
``main'' the role as those who played at least 50 games in the time
period studied, which is roughly an average of two games every three
days.

Similarly, a different set of champions is typically strong in each role
and therefore more likely to be played. For each role, we consider only
games with champions played by at least 30 separate accounts.

The number of games required to include an account in the data and the
number of players required to include a champion in the data were chosen
based on simulations. The simulations used the data availability pattern
of high-ranking accounts in NA on patches 13.6 - 13.9, which were then
filtered based on different requirements. An intercept, an account
random effect, a champion random effect, and random noise were simulated
to generate fake data. A simple random effects model was then fit to the
fake data and the estimated random effects were compared to the true
random effect sizes. One hundred iterations of each filtering pattern
were simulated and then fit in this manner. As we are interested in
whether the estimated player random effect is ordered in the same way as
the actual player effect, the final set of requirements was chosen based
on the method which had the highest correlation between the estimated
and true effect sizes while maintaining a reasonable data set size.

\hypertarget{summary-of-data}{%
\subsection{Summary of data}\label{summary-of-data}}

Table~\ref{tbl-data-summary} provides summary statistics of our data,
after filtering based on the requirements described above. The data are
stratified by server (NA, EUW, and KR) and by role (top, jungle, mid,
bot, and support).

As of patch 13.18, there were 165 total possible champions in the game.

\hypertarget{tbl-data-summary}{}
\begin{longtable}{l|rrrrr}
\caption{\label{tbl-data-summary}Summary statistics of final data set. Models are fit separately to each
region and role subset. }\tabularnewline

\toprule
\multicolumn{1}{l}{} & Top & Jungle & Mid & Bot & Support \\ 
\midrule\addlinespace[2.5pt]
\multicolumn{6}{l}{N} \\ 
\midrule\addlinespace[2.5pt]
EUW & $54,047$ & $59,406$ & $56,888$ & $61,029$ & $57,674$ \\ 
KR & $63,704$ & $67,746$ & $64,780$ & $72,434$ & $64,810$ \\ 
NA & $43,719$ & $51,211$ & $41,578$ & $48,253$ & $45,339$ \\ 
\midrule\addlinespace[2.5pt]
\multicolumn{6}{l}{\# games} \\ 
\midrule\addlinespace[2.5pt]
EUW & $40,529$ & $43,329$ & $41,806$ & $44,528$ & $42,257$ \\ 
KR & $48,004$ & $49,951$ & $49,029$ & $52,361$ & $48,163$ \\ 
NA & $34,664$ & $39,549$ & $33,344$ & $37,797$ & $35,525$ \\ 
\midrule\addlinespace[2.5pt]
\multicolumn{6}{l}{\# champions} \\ 
\midrule\addlinespace[2.5pt]
EUW & $46$ & $44$ & $51$ & $28$ & $37$ \\ 
KR & $41$ & $36$ & $43$ & $24$ & $36$ \\ 
NA & $43$ & $42$ & $47$ & $26$ & $35$ \\ 
\midrule\addlinespace[2.5pt]
\multicolumn{6}{l}{\# accounts} \\ 
\midrule\addlinespace[2.5pt]
EUW & $396$ & $432$ & $426$ & $449$ & $424$ \\ 
KR & $434$ & $480$ & $465$ & $506$ & $474$ \\ 
NA & $386$ & $434$ & $390$ & $437$ & $405$ \\ 
\bottomrule
\end{longtable}

\hypertarget{game-data}{%
\subsection{Game data}\label{game-data}}

The Riot API provides snapshots of the game state at one-minute
intervals. This includes statistics like the player's location on the
map, the amount of gold they have earned, their champion's level, and so
on. We focus on statistics at 7 minutes, 15 minutes, and 25 minutes. 7
minutes and 15 minutes roughly correspond to the end of the early and
mid-game phases. Roughly 50\% of games last past 25 minutes, so
examining the statistics at 25 minutes gives sufficient time to capture
the post-laning phase while maintaining a reasonable dataset size
(20,612, 16,896, and 17,769 in our data for NA, EUW, and KR,
respectively).

We use the ``damage done to champions'', ``damage taken'', and ``total
gold'' statistics provided by Riot's snapshots API. Damage done to
champions is the total damage a player has dealt to any other player in
the game, after accounting for the effects of armor and magic resist.
Damage taken is the total damage a player has taken, after accounting
for \emph{armor} and \emph{magic resist}. Total gold is the total gold
accrued by that time point.

Note that these statistics are cumulative---for example, total gold at
15 minutes includes gold accrued from 0-7 minutes. We partition the
statistics by considering only the change in value within each phase.
So, when considering gold at 25 minutes, we subtract total gold at 15
minutes from total gold at 25 minutes to get the gold accrued during the
15 - 25 minute time period.

All statistics are centered and scaled. All statistics are non-negative
and exhibit slight skew, but as log and square root transforms of the
statistics showed no improvement, no further transformations were
applied.

\hypertarget{sec-linking}{%
\subsection{Linking Riot accounts to professional
players}\label{sec-linking}}

To help validate our methods, we compare model results for known
professional players to all other accounts included in our data. We
define a professional player as someone who plays for a team in the
top-level professional league in North America, Europe, Korea, or China.
The respective top-level leagues for each region are the League
Championship Series (LCS), League of Legends European Championship
(LEC), League of Legends Champions Korea (LCK), and the League of
Legends Pro League (LPL). We include players from the LPL because many
LPL players have accounts on the Korean server.

Professional player accounts were identified by gathering summoner names
from three websites: https://www.trackingthepros.com/, https://gol.gg/,
https://www.deeplol.gg, and https://lolpros.gg/.

Table~\ref{tbl-linking} gives the number of known player accounts found
within our data.

\hypertarget{tbl-linking}{}
\begin{longtable}{crrrrrr}
\caption{\label{tbl-linking}Summary of the number of player accounts within our data which were
linked to a pro player within the LCS, LEC, LCK, or LPL. }\tabularnewline

\toprule
 & \multicolumn{2}{c}{EUW} & \multicolumn{2}{c}{KR} & \multicolumn{2}{c}{NA} \\ 
\cmidrule(lr){2-3} \cmidrule(lr){4-5} \cmidrule(lr){6-7}
Role & \# accounts & \# players & \# accounts & \# players & \# accounts & \# players \\ 
\midrule\addlinespace[2.5pt]
Top & 18 & 15 & 13 & 13 & 10 & 10 \\ 
Jungle & 22 & 21 & 19 & 19 & 4 & 4 \\ 
Mid & 16 & 15 & 18 & 18 & 5 & 5 \\ 
Bot & 25 & 22 & 17 & 17 & 9 & 8 \\ 
Support & 12 & 10 & 16 & 15 & 12 & 9 \\ 
\bottomrule
\end{longtable}

Note that this is not an exhaustive list and many professional players
have alternate accounts that they keep secret.

\hypertarget{methods}{%
\section{Methods}\label{methods}}

League of Legends is an invasion game where the goal is to destroy the
enemy base. As such, progress is often measured by how much of the map a
team can control. If the team can continuously advance towards the enemy
base, then they are ahead. Teams establish this control by taking
objectives, mainly through destroying enemy towers which provide both
vision and defense to the enemy.

Advancing into enemy territory typically requires a balance of two
umbrella categories of skill: ability to gain resources and skill in
using those resources. These categories can be further broken down into
sub-categories such as strategy, tactics, mechanics, and technical
skill, but the end-goal is always to either gain resources or to use
them effectively. We thus focus on gold (resource gain) and damage dealt
(resource use) as our two main statistics. Further explanation on why we
focus on gold and damage dealt specifically is included in
Section~\ref{sec-goldisskill}.

League of Legends is a cooperative game and a player's gold and damage
dealt statistics are rarely independent from their allies' or their
enemies' statistics. For example, some junglers may frequently roam to
top lane to ambush the enemy top laner, giving their top laner the kills
and the resultant gold. Without the jungler's assistance, the gold from
the kills would not have gone to the top laner. These are strategic
decisions which reflect skill on the jungler's part that would go
unnoticed if looking at the jungler's statistics alone. Therefore, we
consider not just a player's own gold and damage dealt statistics, but
also evaluate how much of their allies' and enemies' numbers can be
attributed to the player. We partition a player's gold or damage dealt
into three components: the portion that the player attained for
themselves, the portion they enabled in their allies, and the portion
they prevented from their enemies.

This is difficult to do in a competitive setting, but simpler to do with
solo queue data. In a competitive setting, a player's effect on their
own statistics and their teammate's are highly confounded---and if the
team keeps the same roster the entire season, then they are perfectly
confounded. However, the semi-random matchmaking environment of solo
queue means that teammates are rarely constant from game to game, which
makes it possible to separate the player's effects from their allies'
effects.

Let \(P\) be the set of all players, \(G_p\) be the set of games that
include player \(p\) for \(p \in P\), and \(C\) be the set of champions.
For \(c \in C\), \(g \in G_p\), we can summarize player \(p\)'s effect
on their games by:

\begin{enumerate}
\def\labelenumi{\arabic{enumi}.}
\item
  Estimating player \(p\)'s effect on their own statistics,
  \(\text{gold}_{pg}\) and \(\text{dmg}_{pg}\) (Sections
  \ref{sec-model-gold} and \ref{sec-model-dmg}).
\item
  Estimating player \(p\)'s effect on their allies' statistics
  (Section~\ref{sec-model-ally}). To do this, we first remove the
  portion of the allies' statistics that are not due to player \(p\),
  and then estimate player \(p\)'s effect using the remainder.

  \begin{enumerate}
  \def\labelenumii{\alph{enumii}.}
  \item
    Let \(A_{g}\) be the set of player \(p\)'s allies in game \(g\). For
    each ally \(a\) in \(A_{g}\), first estimate the effect of the ally
    on their own statistics, \(E[\text{gold}_{ag}]\) and
    \(E[\text{dmg}_{ag}]\).
  \item
    Estimate the effect of player \(p\) on their allies using the
    residuals \(\text{gold}_{ag} - E[\text{gold}_{ag}]\) and
    \(\text{dmg}_{ag} - E[\text{dmg}_{ag}]\), summed over all allies
    \(a\), as the model response.
  \end{enumerate}
\item
  Estimating player \(p\)'s effect on their enemies' statistics
  (Section~\ref{sec-model-ally}). To do this, we first remove the
  portion of the enemies' statistics that are not due to player \(p\),
  and then estimate player \(p\)'s effect using the remainder.

  \begin{enumerate}
  \def\labelenumii{\alph{enumii}.}
  \item
    Let \(E_{g}\) be the set of player \(p\)'s enemies in game \(g\).
    For each enemy \(e\) in \(E_{g}\), estimate the effect of the enemy
    on their own statistics, \(E[\text{gold}_{eg}]\) and
    \(E[\text{dmg}_{eg}]\).
  \item
    Estimate the effect of player \(p\) on the enemy team using the
    residuals \(\text{gold}_{eg} - E[\text{gold}_{eg}]\) and
    \(\text{dmg}_{eg} - E[\text{dmg}_{eg}]\), summed over all enemies
    \(e\), as the model response.
  \end{enumerate}
\end{enumerate}

This approach is motivated by being able to capture, at least in part,
subcategories of skill that typically are not quantified. Mechanics and
technical skills involve things like aim, reaction time, last-hitting,
and teamfighting which are often reflected in a player's individual
statistics. Strategic skills like map movements, teamfight formations,
and tactical skills like adapting to the enemy team's strategy typically
have broader impacts and so are noticed less unless a player's effect on
their allies or enemies can be quantified. This is not to say that a
player's individual statistics solely reflect mechanics and technical
skill, simply that there are more facets to player skill that have not
been quantified and so are under-discussed.

The following sections detail the player, ally, and enemy models.
Examples of the four sub-categories of skill are in the Appendix,
Section~\ref{sec-skill-categories}.

\hypertarget{sec-goldisskill}{%
\subsection{Gold and damage dealt as measures of player
skill}\label{sec-goldisskill}}

Progress in the game of League of Legends requires a balance of between
skill in gainng resources and skill in using those resources. Gaining
resources allows players to buy power. A gold differential, where one
team is ahead of another, often also means a power differential. This
makes it easier to force an enemy player off the map by dealing enough
damage to them so that they either die or are forced to retreat. With a
numbers advantage, objectives like enemy towers become much easier to
secure.

\begin{figure}[!ht]

{\centering \includegraphics{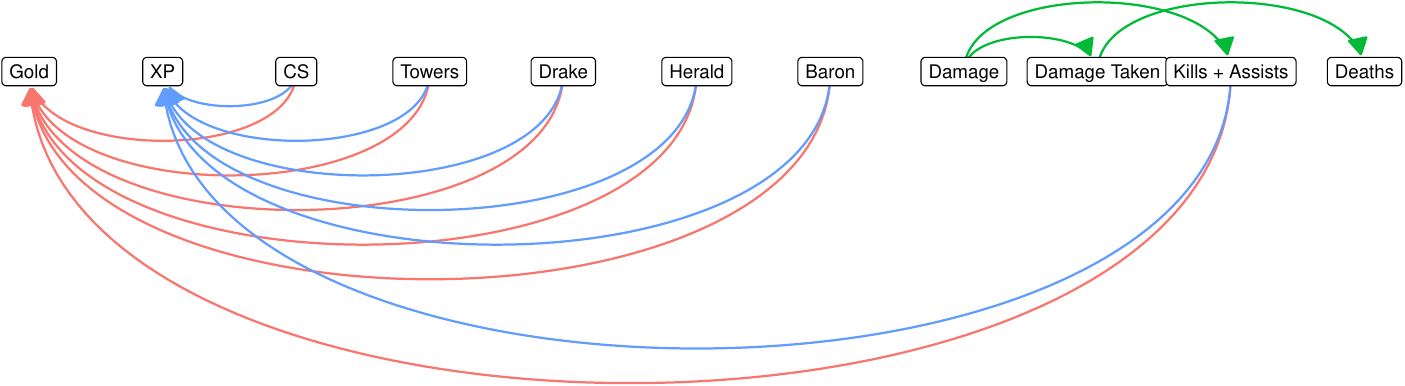}

}

\caption{\label{fig-gold-map}Directed graph plot of relationships
between statistics in the Riot API.}

\end{figure}

As a metric, gold is both meaningful and interpretable. It is commonly
accepted within the community as a measure of how ahead a player or team
is. Player and team gold levels are central statistics displayed and
discussed during Riot broadcasts of professional games. This is both
because gold is a rough measure of power in a game and also because most
measurable and meaningful actions within the game result in a gold gain.
Figure~\ref{fig-gold-map} is a relationship of statistics provided by
the Riot API. Two statistics are connected if gain in one statistic
typically causes a gain in the other. Note that the majority of
statistics are directly connected to gold. Those that aren't, are
indirectly correlated via other statistics.

\begin{figure}[!ht]

{\centering \includegraphics{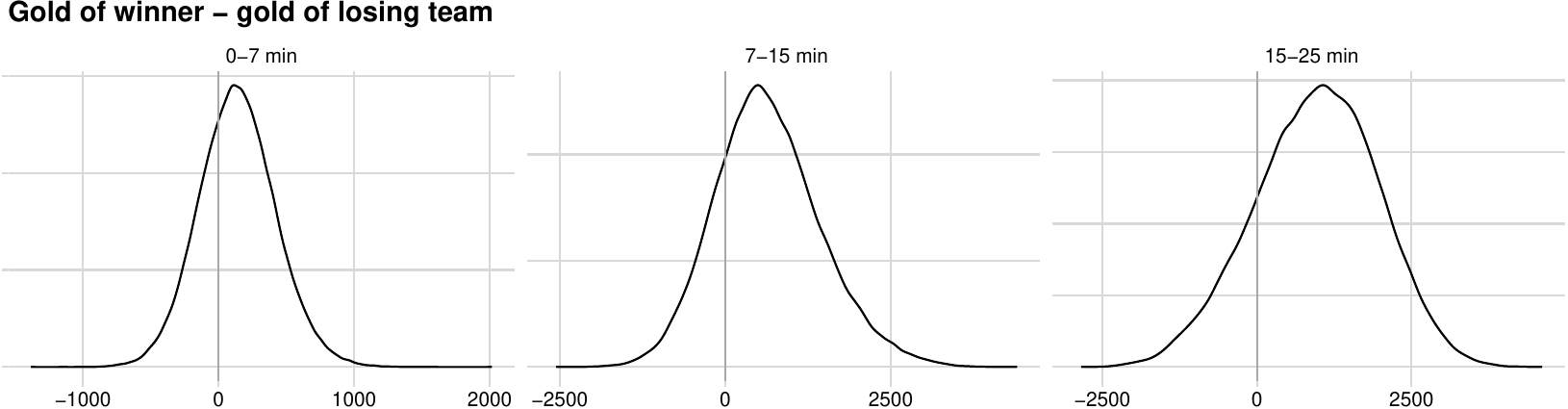}

}

\caption{\label{fig-gold-differential}Density plots of gold differential
between the winning and losing team in our data set. Percent of games
with gold differential above 0 is 69\% at 0-7 minutes, 79\% at 7-15
minutes, and 83\% at 15-25 minutes.}

\end{figure}

Gold is also predictive of who will win.
Figure~\ref{fig-gold-differential} provides density plots of the
difference in gold accrued by the winning team vs the losing team within
each time span. The gold differential is typically positive. The later
in the game, the higher gold leads by the winning team tend to be. Note,
however, that the winning team is not always ahead in gold. This
reflects one characteristic of the game, which is that a good
performance does not necessarily lead to a win, and a win does not
necessarily justify any single player's individual performance.

\begin{figure}[!ht]

{\centering \includegraphics{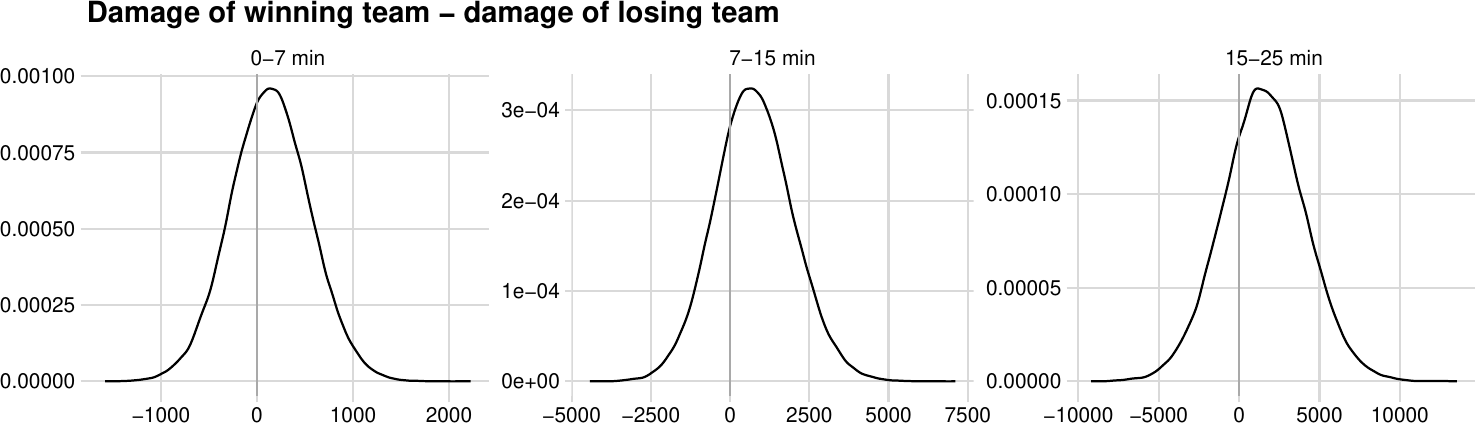}

}

\caption{\label{fig-dmg-differential}Density plots of damage
differential between the winning and losing team in our data set.
Percent of games with gold differential above 0 is 63\% for 0-7 minutes,
73\% for 7-15 minutes, and 72\% for 15-25 minutes.}

\end{figure}

In League of Legends, gold is often seen as a metric for measuring a
player's efficiency in resource collection. However, the true
effectiveness of a player isn't just in the accumulation of resources,
but also in how effectively these resources are utilized in the game.
Fundamentally, League of Legends is a strategic contest between two
teams, each aiming to capture the opponent's base. Conflict is an
intrinsic part of this contest, and the gold and experience players
accumulate translate primarily into combat strength. This strength is
crucial for overpowering the enemy, advancing into their base, and
securing key resources on the map. Similar to gold, damage is associated
with wins, as seen in Figure~\ref{fig-dmg-differential}, with winning
teams dealing more damage in \textasciitilde70\% of cases during later
periods of the game.

More than measuring just combat performance, damage dealt can also serve
as an indication of technical and mechanical skill. For instance, a
player with superior aiming skills is more likely to land skillshots,
resulting in higher damage output over a period. Similarly, a player
with quick reflexes can evade enemy abilities more effectively, stay
engaged in teamfights longer, and consequently deal more damage while
sustaining the same amount of health.

Furthermore, it's important to recognize the roles of players,
champions, and team positions that aren't primarily focused on dealing
damage. Their actions in teamfights, whether it's enabling teammates to
deal damage or hindering the enemy's ability to inflict harm, are
equally pivotal. They contribute to the team's overall combat strategy
and effectiveness. Damage as a metric is particularly insightful because
it demonstrates that a player or team can overcome disadvantages through
superior strategy, better coordination, or simply outclassing the
opposing team in skill. Even when at a numerical or positional
disadvantage, a team can triumph in a fight with the right approach and
execution. Therefore, while damage is not the sole indicator of skill,
it is a significant and measurable aspect of a player's contribution to
the team's success in League of Legends.

\hypertarget{sec-model-gold}{%
\subsection{Player model: Gold}\label{sec-model-gold}}

Regression models allow for separating the effects of a player's skill
on the gold they've accrued from the effects of confounders. Here, the
main confounder we are concerned with is the player's choice of
champion. Different champions are expected to perform better at
different stages of the game, known within the game as ``scaling'', and
so we do not expect average gold or damage dealt to be the same across
champions. Other confounders, such as time effects, were also considered
but no meaningful effects were found.

Let \(P\) be the set of all players, \(G_p\) be the set of games that
include player \(p\) for \(p \in P\), and \(C\) be the set of champions.
For \(c \in C\), we control for the player's choice of champion with a
champion random effect, \(b_c\). We estimate the average impact a player
has on their own statistics with a player-specific random effect,
\(b_p\). For \(g \in G_p\), the regression model used for the gold
statistic is as in Equation~\ref{eq-gold-own}:
\begin{equation}\protect\hypertarget{eq-gold-own}{}{
\text{gold}_{pg} = \beta_0 + b_c + b_p + \epsilon_{pg},
}\label{eq-gold-own}\end{equation} where \(\beta_0\) is an intercept
term, \(\epsilon_{pg}\) is random noise, and \(\sigma^2\) is the
variance for \(\epsilon_{pg}\).

The model is fit in a Bayesian paradigm, which has been shown to perform
well in applications with scarce and/or imbalanced data, a property
which is often attributed to hierarchical shrinking and information
pooling (Gelman 2006; Morris 1983). For this reason, many recent models
in sports analytics take a Bayesian approach (Herbrich, Minka, and
Graepel 2007; Clark, Macdonald, and Kloo 2020; Santos-Fernandez, Wu, and
Mengersen 2019; Mengersen et al. 2016). In a Bayesian paradigm, all
parameters \(\Theta\) have an associated prior probability density
\(f(\Theta)\) which represents reasonable beliefs on the scale of the
parameters. These beliefs are updated by sampling from the conditional
distribution \(f(\Theta | Y)\), which is called the posterior
distribution. Given the posterior distribution, we can then calculate
summary statistics such as \(E[\Theta | Y]\) or even sample from the
posterior predictive distribution \(f(Y_{\text{new}} | Y)\) to make
predictions on new data.

Let \(t_3\) denote a Student's-t distribution with three degrees of
freedom and let \(\text{HalfC}(x)\) denote a half-Cauchy distribution
with scale parameter \(x\). The priors in this model are
\(\beta_0 \sim N(0, 1)\), \(b_c \sim t_3(0, \phi)\),
\(b_p \sim t_3(0, \tau)\), and \(\phi, \tau \sim \text{HalfC}(0.5)\),
\(\sigma \sim \text{HalfC}(0.3)\).

Note here that the random effects have scale hyperparameters \(\tau\)
and \(\phi\); this makes the model a Bayesian hierarchical regression
model. A Student's-t distribution is used instead of a normal density
because it has heavier tails, which allows for outliers in the champion
or player random effects. All scale hyperparameter values (\(\phi\),
\(\tau\), \(\sigma\)) have half-Cauchy priors as the half-Cauchy has
been shown to produce better estimates for scale parameters (Gelman
2006; Polson and Scott 2012). A HalfC(0.5) prior has a median value of
0.5, while a HalfC(0.3) prior has a median value of 0.3. Prior parameter
values are chosen to be weakly informative based on reasonable beliefs
in the size of effects (Gelman, Simpson, and Betancourt 2017); since all
data are centered and scaled, we chose prior parameters that reflect the
belief that variance parameters are between 0 and 1. We also compared
the prior predictive distribution to the density of the data as a sanity
check that the priors were not over-shrunk towards zero.

We use the posterior estimates for the player random effect, \(b_p\), as
our metric for player skill. Similar to a Plus-Minus statistic, this
presents a naturally ordered list where players with large positive
scores are taken as more skilled in gaining gold than average, while
players with negative scores are less skilled. Because the statistics
are centered and scaled, we can interpret \(b_p\) in terms of standard
deviations. A value of \(1\), for example, means that a player's gold
generation is, on average, one standard deviation above the mean,
relative to all games. If divided by \(\tau\), \(b_p / \tau\) can be
interpreted as a score relative to all other players. We use \(b_p\)
directly so that we have a meaningful measure of effect size; a small
\(\tau\) means the player score will also be appropriately small.

\hypertarget{sec-model-dmg}{%
\subsection{Player model: Damage dealt}\label{sec-model-dmg}}

Damage dealt can be high because a player is particularly effective at
dealing damage or simply because they are more aggressive. To assess
skill in dealing damage, then, we must be able to standardize the amount
of damage dealt by the frequency of skirmishes a player is involved in.
Since it is rare for damage to be dealt without also taking damage (and
vice versa), we include a damage taken covariate as a rough estimate of
how aggressive a player is. Otherwise, the model is the same structure
as the gold model.

Let \(P\) be the set of all players, \(G_p\) be the set of games that
include player \(p\) for \(p \in P\), and \(C\) be the set of champions.
For \(c \in C\) and \(g \in G_p\), the model for damage dealt is shown
in Equation~\ref{eq-dmg-own} below.

\begin{equation}\protect\hypertarget{eq-dmg-own}{}{
\text{dmg}_{pg} = \beta_0 + \beta_{\text{dmgt}} x_{pg} + b_c + b_p + \epsilon_{pg}
}\label{eq-dmg-own}\end{equation}

where \(x_{pg}\) is the amount of damage taken by player \(p\) in game
\(g\) during the given time period in the given role. The prior placed
on \(\beta_{\text{dmgt}}\) is \(N(0, 1)\). All other priors are the same
as in the gold model.

\hypertarget{sec-model-ally}{%
\subsection{Ally models}\label{sec-model-ally}}

When estimating a player's impact on their allies' statistics, we first
account for each ally's champion, role, and average gold or damage
accrued and remove that from their gold or damage numbers. In a
Frequentist context and under the semi-random matchmaking environment of
solo queue, these steps may not be necessary so long as a player is not
consistently matched with the same ally playing the same role or the
same champion. In other words, we would not expect any confounding and
therefore would not expect any multicollinearity. However, given the
high degree of data imbalance in the data, the prevalence of one-tricks
(players who play the same champion and role in the vast majority of
their games), and the relatively small pool of high-ranking players who
are all playing at different times, meaning the same players can be
matched together again and again, we do not assume independence.

We determine an ally's expected gold or damage dealt values and then use
the total amount all allies over- or under-perform relative to their
expected quantity as the model response. Specifically:

\begin{enumerate}
\def\labelenumi{\arabic{enumi}.}
\item
  Let \(G_p\) be the set of all games involving player \(p\) and let
  \(g \in G_p\). Let \(A_{g}\) be the set of player \(p\)'s allies in
  game \(g\). For each ally \(a\) in \(A_{g}\), estimate the expected
  gold accrued or damage dealt by ally \(a\), \(E[\text{gold}_{ag}]\)
  and \(E[\text{dmg}_{ag}]\). This is done using the models in
  Equation~\ref{eq-gold-own} and Equation~\ref{eq-dmg-own},
  respectively, but expanded for all allies of players in our data. Note
  that this requires re-fitting the models to the expanded set of data.
\item
  Take the sum total of the residuals within each game,
  \[\Delta(\text{gold}_{A_{g}}) := \sum_{a \in A_{g}} \left(\text{gold}_{ag} - E[\text{gold}_{ag}]\right)\]
  and
  \[\Delta(\text{dmg}_{A_{g}}) := \sum_{a \in A_{g}} \left(\text{dmg}_{ag} - E[\text{dmg}_{ag}]\right)\]
\item
  Fit a mixed-effects regression model to the sum of residuals,
  \(\Delta\), \begin{equation}\protect\hypertarget{eq-allyenemy}{}{
  \Delta = \beta_0 + b_c + b_p + \epsilon_{g}.
  }\label{eq-allyenemy}\end{equation}
\end{enumerate}

We use the posterior estimate for \(b_p\) to assess the average impact
player \(p\) has on their four allies.

By first estimating and then removing a single ally \(a\)'s impact on
their own statistics, we account for the champion and role ally \(a\) is
playing, as well as the amount of gold and damage ally \(a\) typically
accrues within our data. If any multicollinearity is present, this
amounts to saying that any overlap in information should be attributed
to ally \(a\), rather than player \(p\). The remainder of ally \(a\)'s
statistics is then due to the other players in the game and the
particular circumstances of that single game. By fitting the
mixed-effects model in Equation~\ref{eq-allyenemy}, we can determine the
total impact player \(p\) has on their allies as a collective.

We model a player's impact on their allies as a team because we
typically found that, outside of their direct lane opponent, the
majority of players had little impact on their allies in our data,
leading to estimates with small effect sizes and wide posteriors.

Re-fitting the models in step 1 to the larger set of data can be
expensive in terms of computing time. Since including all allies or
enemies increases the sample size of the data, it also increases the
fitting time of the model. To speed up computation time, we approximate
the re-fitted model estimates using AXE (A. X. Zhang et al. 2023). AXE
produces reliable estimates of the fitted values for a re-fitted
Bayesian hierarchical regression model as long as the posterior means of
the variance hyperparameters (in this case, \(\tau\), \(\phi\), and
\(\sigma\)) are similar between the original model and the re-fitted.
Note that this usage is an extension of the AXE method, as we first fit
to the smaller subset and then use AXE to approximate posterior means
for the model fitted to the larger data, rather than the reverse.

\hypertarget{sec-model-enemy}{%
\subsection{Enemy models}\label{sec-model-enemy}}

The procedure for the enemy models is the same as the ally model, using
gold/damage values for the set of all enemies \(e\) in \(E_{g}\). Rather
than the posterior estimate for \(b_p\), we use \(-b_p\) as the player's
performance metric, because reducing the enemies' gold income or damage
dealt is a positive indication of skill for player \(p\).

\hypertarget{champion-proficiency-heuristic}{%
\subsection{Champion proficiency
heuristic}\label{champion-proficiency-heuristic}}

The SIDO models give a score relative to all players, regardless of the
champions they played. An additional heuristic that may be useful is one
that measures a player's skill with a specific champion, relative to all
other players who play the same champion.

Using the SIDO models, the posterior expectation of an average player's
performance on champion \(c\) is \(\hat{\beta}_0 + \hat{b}_c\), where
\(\hat{\beta}_0 := E[\beta_0 | Y]\) and \(\hat{b}_c := E[b_c | Y]\),
where \(Y\) is the model response (own gold or damage dealt, in this
case). Let \(G_{pc}\) be the set of all games player \(p\) plays on
champion \(c\) and \(n_{pc}\) be the cardinality of the set \(G_{pc}\).
To get the player's average performance on a champion, we subtract this
expectation from their actual gold (or damage) values and take the
average of the differences, as in Equation~\ref{eq-champ-heuristic}
below:

\begin{equation}\protect\hypertarget{eq-champ-heuristic}{}{
\delta_{pc} := \frac{1}{n_{pc}}\sum_{g \in G_{pc}}\left(Y_{pg} - \hat{\beta}_0 + \hat{b}_c\right),
}\label{eq-champ-heuristic}\end{equation}

We then center and scale the set of \(\delta_{pc}\) across all players
\(p\) to get the champion proficiency heuristic.

\hypertarget{results}{%
\section{Results}\label{results}}

We compare the SIDO models to two other models: a basic average model
(BA), representing the industry standard, and the Plus-Minus model as
described by Clark, Macdonald, and Kloo (2020). For the BA model, we
simply calculate the average gold (or damage dealt) for each player
within the games in our data set.

\hypertarget{sec-proplayers}{%
\subsection{Separation of pro players}\label{sec-proplayers}}

A difficulty of performance models in general is that there is no
objective truth standard. One way to indirectly validate our results is
by comparing the average player effect size for accounts known to belong
to a professional player to all other accounts. In general, we expect
accounts belonging to professional players to score higher in our model
than the average grandmaster solo queue player.

We compare scores for accounts belonging to pro players to unknown
accounts, i.e.~accounts not linked to a current pro, a retired pro, or
anyone playing in a professional circuit (such as academy and amateur
leagues). We label these unknown accounts as ``non-pro'', though it is
likely they also contain accounts for pros.

For each of the 0-7 minute, 7-15 minute, and 15-25 minute time periods,
we summarize pro scores into two numbers: the ``ally + player'' score,
the sum total of the ally and player models, and the ``enemy'' score.
The BA model does not include ally/enemy scores, so we present only one
score. The Plus-Minus model includes defensive and offensive scores---we
label the defensive scores as ``enemy'' scores and offensive scores as
``ally'' scores.

Overall, we found that there was insufficient evidence that the
Plus-Minus model was able to meaningfully differentiate between pro and
non-pro accounts. The Plus-Minus model often scored pro accounts lower
than the average non-pro account. Most differences were also small in
magnitude. Both the BA and SIDO models found large positive differences
between the average pro and average non-pro scores in top, mid, and bot,
but the BA model showed fewer differences in the jungle and support
roles, which are often less carry-focused and more utility-focused. The
SIDO model found large positive differences across all roles.

Plots describing our results follow in the next sections.

\hypertarget{gold-models}{%
\subsubsection{Gold models}\label{gold-models}}

\begin{figure}[!ht]

{\centering \includegraphics{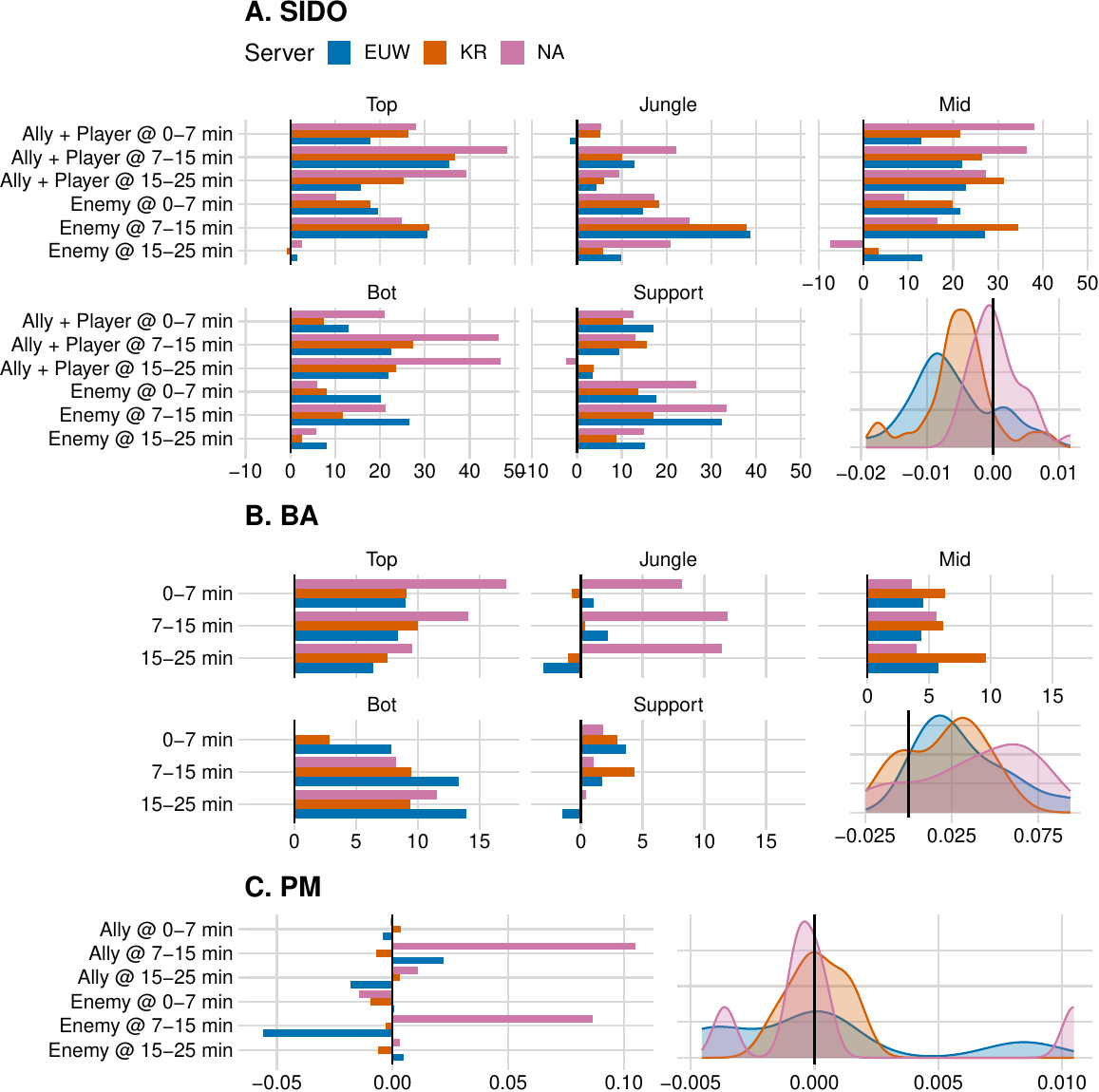}

}

\caption{\label{fig-pros-gold}For all damage models, bar charts of the
average difference between pro and non-pro scores and density plots of
average non-pro scores. Differences are divided by the standard
deviation of non-pro scores, so that different methods may be compared.}

\end{figure}

Figure~\ref{fig-pros-gold} provides bar charts of the differences
between pro and non-pro accounts for the SIDO, BA, and Plus-Minus gold
models, respectively.

We see a considerable positive difference between pro and non-pro scores
in almost all regions, game-segments and roles under the SIDO models,
with the exception of ``Enemy @ 15-25 min'' for top, mid and bot, and
``Ally + Player @ 15-25'' for support. These may be a result of the
changing priorities or opportunities during that phase of the game. For
example, during that time period the laners (top, mid, adc) often work
together to synchronize their movement to set up for and capture
important objectives instead of prioritizing the denial of resources
from their opponents. Similarly, the 0-7 min time period for junglers
involves very few opportunities for differentiation as they spend most
of their time collecting resources and looking for ways to capitalize on
their counterpart's inefficiencies or enemy weaknesses rather than help
allies.

Although the BA model also shows some differences between pro and
non-pro scores, it is less consistent than the SIDO model, and the
magnitude of such differences are smaller. For most regions, the BA
model does not distinguish between pro and non-pros in the jungle and
support roles. For the mid and bot roles, it's interesting to see an
increase of differences over the course of the game, with the opposite
true for top role---this is consistent with late-game differences we've
seen for the SIDO model.

The Plus-Minus model overall shows much smaller differences between pro
and non-pro accounts for both ``ally'' and ``enemy'' models. The density
plot of average scores also reveal some scores for NA which might
account for the larger differences for ``Ally @ 7-15 min'' and ``Enemy @
7-15 min'', but in terms of magnitude expressed per standard deviation,
these still fall considerably short of the SIDO or BA models.

\hypertarget{damage-models}{%
\subsubsection{Damage models}\label{damage-models}}

\begin{figure}[!ht]

{\centering \includegraphics{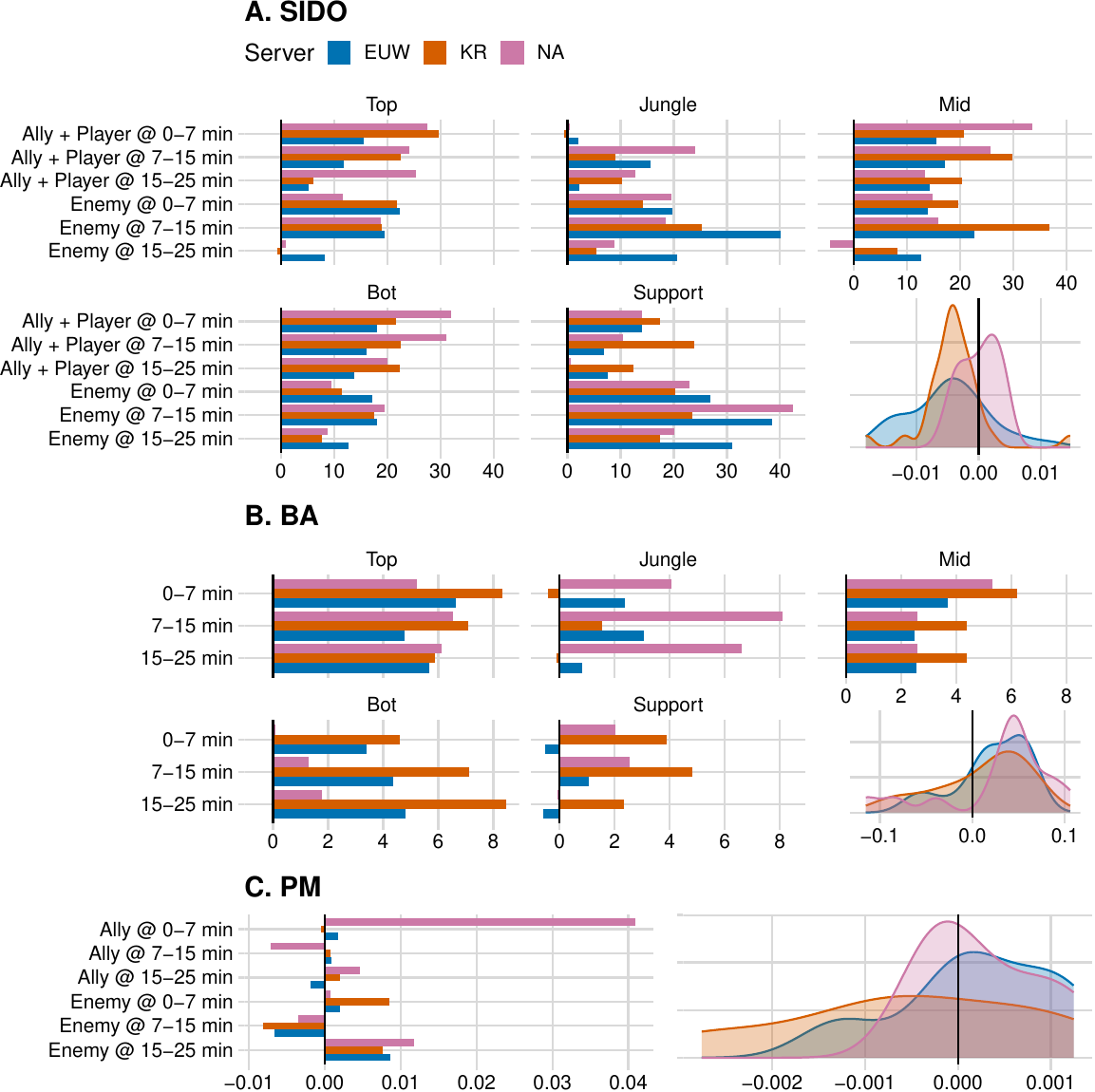}

}

\caption{\label{fig-pros-dmg}For all gold models, bar charts of the
average difference between pro and non-pro scores and density plots of
average non-pro scores. Differences are divided by the standard
deviation of non-pro scores, so that different methods may be compared.}

\end{figure}

Figure~\ref{fig-pros-dmg} provides bar charts of the differences between
pro and non-pro accounts for the SIDO, BA, and Plus-Minus damage models,
respectively.

The SIDO models show consistent differences between pro and non-pro
players. Notably the support role, whose value is traditionally hard to
quantify, shows a significant differentiation through the enemy damage
prevented throughout the game. This would highlight that for support
players, professionals are significantly better at understanding the
enemy players and preventing their damage. The lack of difference in
some of the stats may result in specific role priorities. For example,
the first seven minutes for junglers involves very few opportunities for
differentiation due to the nature of their role. A similar reasoning
regarding their approach to the role in teamfights and skirmishes for
mid and top laners after 15 min could explain the lack of difference in
the enemy damage prevented.

The BA damage models look considerably similar to the BA gold models,
which showcases the relationship between gold (resource) and damage
(power) that we highlighted earlier. We note that the regional
differences in the BA damage models are larger than for gold, especially
for mid, bot and support roles.

Similar to the gold scores, the Plus-Minus model shows considerably
smaller differences and less consistency, making it unable to
distinguish between pro and non-pro players. We will examine the model
in more detail in section 7.2.

\hypertarget{sec-ttests}{%
\subsubsection{SIDO model t-tests}\label{sec-ttests}}

\begin{figure}[!ht]

{\centering \includegraphics{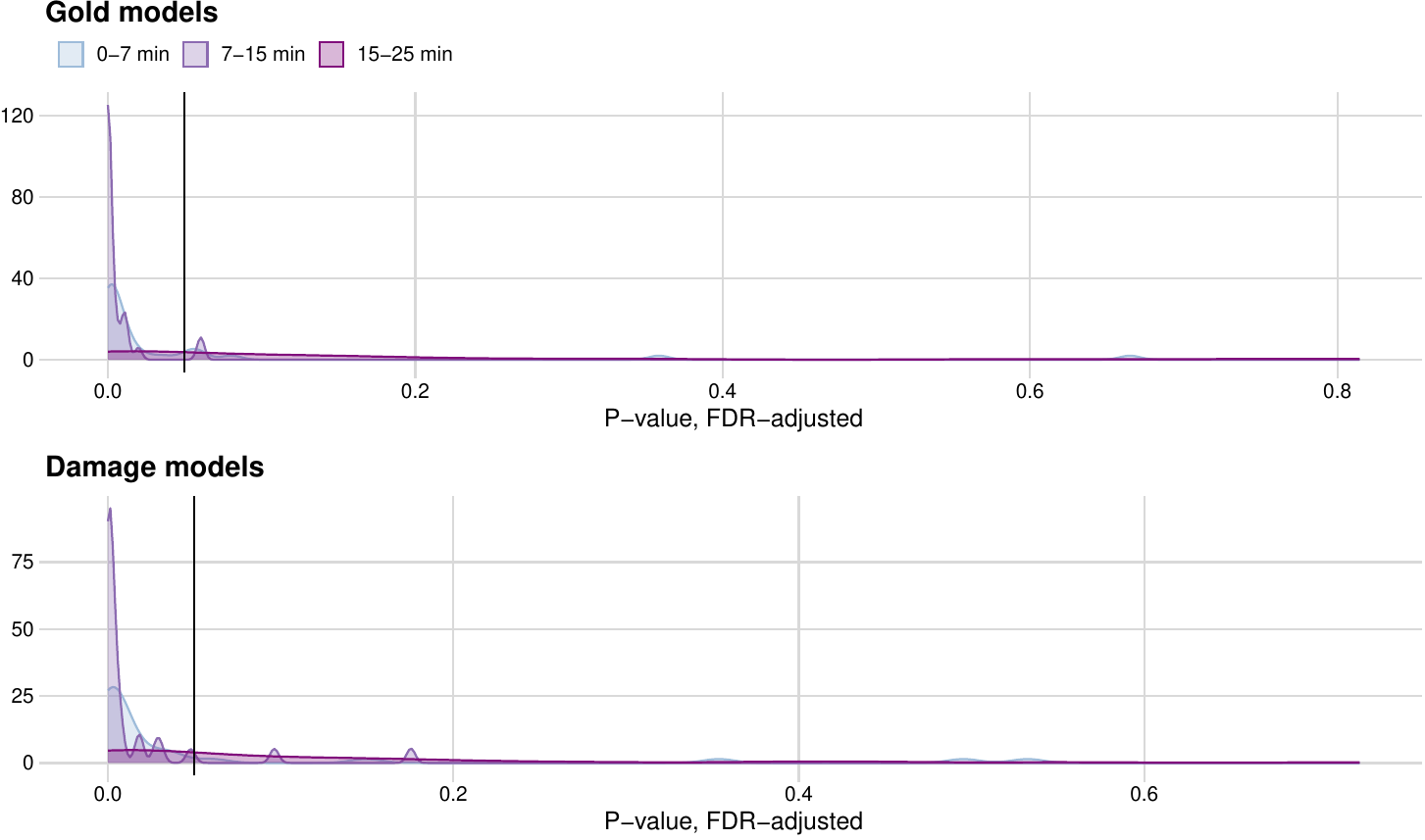}

}

\caption{\label{fig-tests-gold}Plots of FDR-adjusted p-values for all
SIDO models. Vertical line is present at p-value of 0.05.}

\end{figure}

Subfigures \ref{fig-pros-gold}A and \ref{fig-pros-dmg}A show many large
positive differences between pro and non-pro scores under the SIDO
model. We conducted one-sided t-tests of these differences to determine
which were statistically significant under the alternative hypothesis
that the average pro score is greater than the average non-pro score.
Correction for multiple testing was applied using Benjamini-Hochberg's
false discovery rate adjustment (FDR, Benjamini and Hochberg 1995),
which was shown to have more power than the Bonferroni correction.
Figure~\ref{fig-tests-gold} presents density plots of the FDR-adjusted
p-values. The majority of the differences in the 0-7 minute and 7-15
minute time period are large enough to be statistically significant. All
of the negative differences have p-values larger than 0.5, indicating
that the difference is too small to reject the null. For the 15-25
minute period, the density of the adjusted p-values is much broader and
in some cases resembles a uniform distribution. Less difference, then,
is detected by the SIDO models between pros and non-pros in the 15-25
minute time period.

\hypertarget{sec-metametrics}{%
\subsection{Metric quality}\label{sec-metametrics}}

Performance metric quality can be assessed using the meta-metrics
defined by Franks et al. (2016), who proposed three separate criteria:

\begin{enumerate}
\def\labelenumi{\arabic{enumi})}
\item
  discrimination: does the metric meaningfully differentiate between
  players?
\item
  independence: does the metric provide new information that other
  metrics do not?
\item
  stability: does the metric measure the same thing over time?
\end{enumerate}

We evaluate the BA model, the Plus-Minus model of Clark, Macdonald, and
Kloo (2020), and our models under the three criteria above.

\hypertarget{sec-discrimination}{%
\subsubsection{Discrimination}\label{sec-discrimination}}

The discrimination meta-metric is the fraction (between 0 and 1) of
between-player variance in a metric that is not attributed to sampling
variance. Thus, metrics which are low in noise tend to score high in
discrimination.

\begin{figure}[!ht]

{\centering \includegraphics{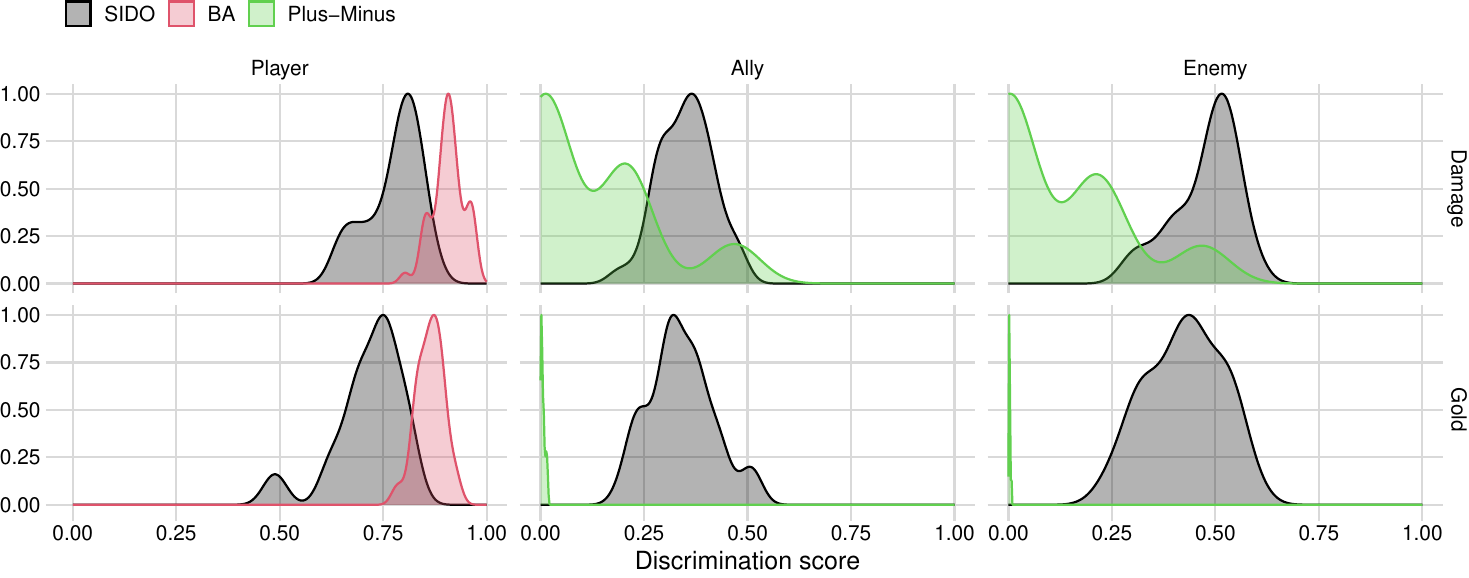}

}

\caption{\label{fig-discrimination}Density plots of discrimination
values for all models.}

\end{figure}

First, we note that the SIDO player models have lower discrimination
than the BA model. This is to be expected, based on the model structure.
The SIDO models separate the response into the portion due to the player
(\(b_p\)) and the portion due to the champion (\(b_c\)), but the BA
model does not take into account champion. Players with different
champion pools will then be scored on uneven scales, creating larger
differentiation. Another reason for lower discrimination is because
hierarchical Bayesian models are inherently more conservative---they
encode more sources of uncertainty and, for those player accounts with
fewer games played, will shrink model estimates for the account towards
a global mean. While we use bootstrapped standard error estimates for
calculating discrimination in the BA model, in the SIDO models we use
the posterior standard deviations. Thus a hierarchical Bayesian version
of the BA model (the same as the SIDO models but without adjusting for
confounders) would also be expected to have lower discrimination scores.

\begin{figure}[!ht]

{\centering \includegraphics{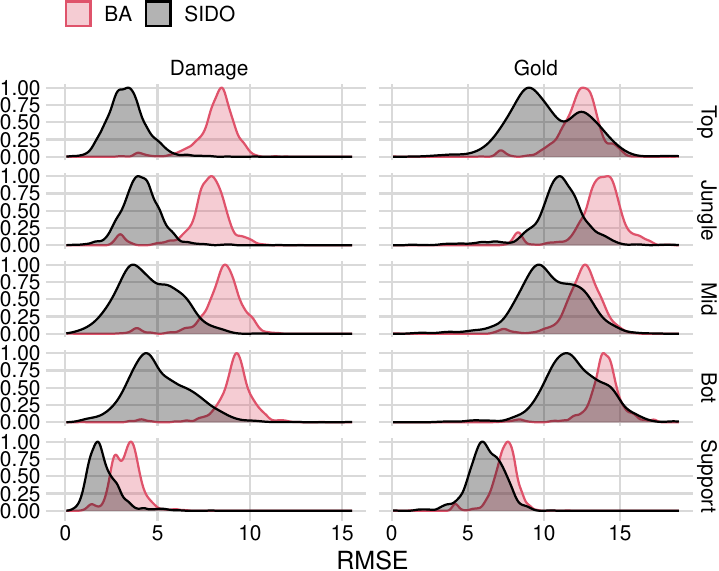}

}

\caption{\label{fig-predictions}Density plots comparing SIDO player
model predictions on test data to BA model predictions.}

\end{figure}

The question, then, is whether the shrinkage and additional uncertainty
encoded in the Bayesian model is appropriate or not. To assess this, we
used both the SIDO and BA models to predict player gold and damage
scores in patches 13.10 and 13.11. We then calculated root mean square
error (RMSE) for each player account. Figure~\ref{fig-predictions} shows
that the SIDO model consistently performs better at prediction, with
lower RMSEs. This indicates that, while the SIDO model performs lower on
discrimination than the BA model, it is more accurate.

Next, we note that discrimination scores for the ally/enemy models are
low across the board. Scoring high in discrimination requires not just
being able to separate the extremes from the remaining accounts, but
also the ability to separate any given pair of accounts from each other.
The results in Section~\ref{sec-proplayers} suggest that the SIDO
ally/enemy models may be able to separate players at the extremes from
the remaining player accounts, but the discrimination results above show
that the ally/enemy model scores contain a good deal of uncertainty that
makes it difficult to distinguish accounts in the middle of the pack
from each other. Rather than providing a continuous score for the
ally/enemy models as we do for the player models, we thus recommend
categorizing ally/enemy scores as ``high positive impact'', ``low
positive impact'', ``neutral'', ``low negative impact'', and ``high
negative impact''.

It is possible that most pro players approach solo queue with a less
cooperative mindset than professional play. We note that discrimination
of the champion effect in the ally/enemy models is more reasonable with
median values of 0.73 and 0.76, respectively. This indicates that effect
sizes may simply be smaller for the ally/enemy models than for the
player models. To verify this, the models need to be extended to pro
play, which we leave for future work.

The SIDO ally/enemy models do out-perform the Plus-Minus models in
discrimination. We posit a few reasons for this: 1) The Plus-Minus
models do not take into account role, champion, or other confounding
variables. For example, a player who mains top lane will have very
different numbers in a jungle role than their main role. All of the
above would lead to greater uncertainty in the model and thus a lower
discrimination score. 2) The Plus-Minus models require fitting a
separate effect for all players in all games. This creates a highly
skewed distribution of player effects, where most player effects belong
to people that are filtered out of the SIDO models as not having enough
high-Elo games. The vast majority of player effects under the Plus-Minus
models are very close to zero, which one would expect under a model
where the majority of players are included in only a handful of games.

The discrimination results indicate that it is necessary to carefully
consider how SIDO model results are presented so that the level of
uncertainty does not lead to misleading conclusions. Part of this is
mitigated by the structure of Bayesian hierarchical models---players
whose numbers carry a large degree of uncertainty will have their
numbers shrunk towards zero. Another method is to save all trace results
from the model so that when any two player numbers are compared to each
other, the posterior probability that one player has a higher score than
another can be computed.

A full table of the discrimination scores is provided in the Appendix,
Section~\ref{sec-app-disc}.

\hypertarget{sec-independence}{%
\subsubsection{Independence}\label{sec-independence}}

Franks et al. (2016) define the independence metric as the fraction of
variance in a metric that is not correlated with any other performance
metric. When independence is high (1), then a metric is independent from
all other performance metrics. The independence metric is calculated by
first estimating correlation using a Gaussian copula model. The
properties of a multivariate normal distribution can then be used to
calculate independence. Note that, based on this definition,
independence tends to decrease as more metrics are compared to each
other, and so the value of the independence metric depends on the set of
metrics that are being considered.

A full table of the independence scores is provided in the Appendix,
Section~\ref{sec-app-ind}.

\begin{figure}[!ht]

{\centering \includegraphics{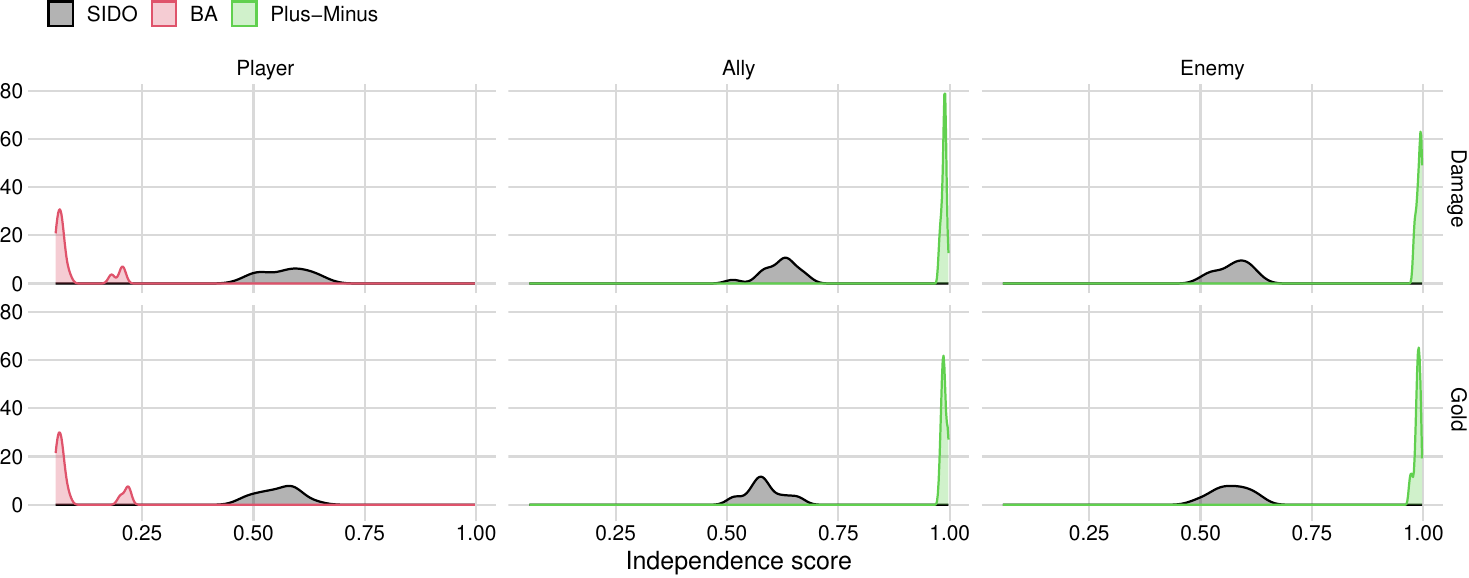}

}

\caption{\label{fig-ind}Density plots of the independence metric across
all regions, roles, and timepoints.}

\end{figure}

The Plus-Minus model scores the highest in independence, while the BA
model scores the lowest. The low scores for the BA model are likely due
to the high correlation between raw gold and raw damage numbers. Without
accounting for champion and other confounders, the raw damage statistic
essentially provides very little new information from the raw gold
statistic. The SIDO models perform in the middle, with numbers generally
between 0.5 and 0.7.

\hypertarget{sec-stability}{%
\subsubsection{Stability}\label{sec-stability}}

Measuring stability over time provides assurance that a metric is
measuring the same thing from patch to patch. However, we note that part
of the usefulness of a good metric is tracking progression of a player's
career over time. Thus, we are interested in determining whether the
SIDO models are relatively stable---meaning that a player who scores
high one patch does not have an extremely low score the next patch---but
they should also reflect any change in ability over large periods of
time.

To assess stability of each method's performance metrics over time, we
fit the SIDO and BA models to data from patches 13.14 - 13.18 and
compared to the metrics from patches 13.6 - 13.9. This corresponds to
dates March 21, 2023 to May 18, 2023. The amount of overlap in player
accounts between the two data sets is around 20\% across all regions, as
shown in Table~\ref{tbl-stab-overlap}. In contrast, the amount of
overlap between the two data sets in champions played in each role is
much higher at around 80\% in most roles and around 90\% in the bot and
top roles. This is likely due to our data collection method, which
selects only games played by the top 1000 players for every day, then
further filters down players based on the number of games. For players
who hover around the boundary for grandmaster, this two-step process
greatly increases the number of games they need to play to be included
in our data.

\hypertarget{tbl-stab-overlap}{}
\begin{longtable}{lrr}
\caption{\label{tbl-stab-overlap}Percent of overlap between the two time periods of data. }\tabularnewline

\toprule
Role & Accounts & Champions \\ 
\midrule\addlinespace[2.5pt]
BOT & 0.17 & 0.90 \\ 
JGL & 0.18 & 0.79 \\ 
MID & 0.19 & 0.77 \\ 
SUP & 0.17 & 0.78 \\ 
TOP & 0.22 & 0.92 \\ 
\bottomrule
\end{longtable}

\newpage{}

We calculated the concordance index between player scores from patches
13.14 - 18 and 13.6 - 13.9. Given the set of all possible pairs of
players within our data, the concordance index counts the number of
pairs which are identically ordered between the two data sets, e.g.~a
concordance index of 0.6 means that 60\% of all possible combinations of
two players have the same ordering between our two data sets. As the
purpose of a performance metric is to compare players to each other, the
concordance index provides an intuitive measure of how similar any two
sets of player rankings are to each other.

For the ally/enemy models, we use the categorizations described in
Section~\ref{sec-discrimination}. We assign number rankings of 1-5 to
the categorizations, which are then used to calculate concordance.
Concordance is scale-free, so the actual numbers assigned to the
categorizations do not matter as long as they are in order.

Given the Plus-Minus model's low discrimination and high computing time
(22 - 96 hours for a single model), we did not re-fit the Plus-Minus
model to the new data set.

\begin{figure}[!ht]

{\centering \includegraphics{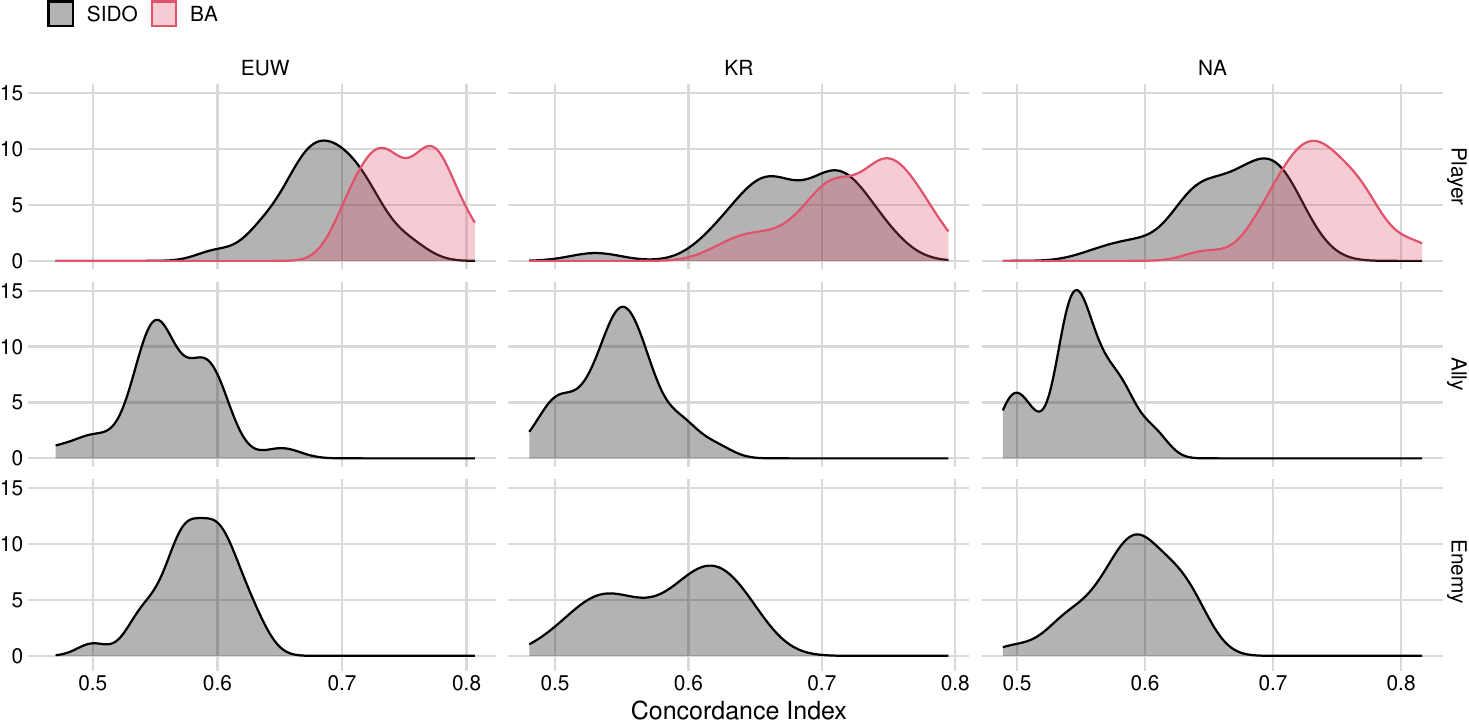}

}

\caption{\label{fig-stability}Density plots of concordance index for
each model and method.}

\end{figure}

\begin{figure}[!ht]

{\centering \includegraphics{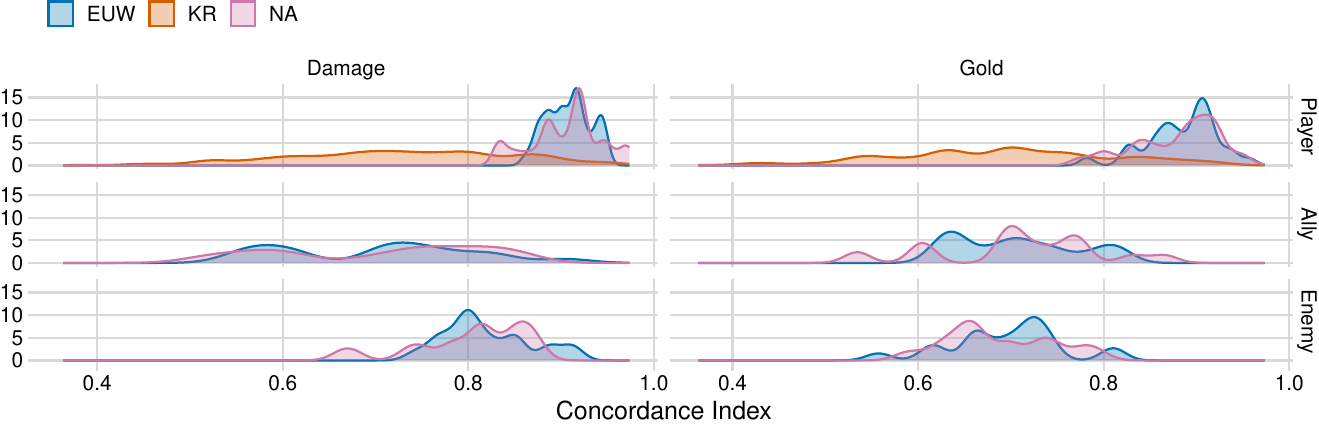}

}

\caption{\label{fig-stab-champion}Density plots of concordance index
values for SIDO ``player'' models, champion effect.}

\end{figure}

Conceptually, the player SIDO models are similar to the BA model, except
they account for the champion played and include additional shrinkage
based on the structure of the data. In other words, the BA model, to
some degree, conflates the player scores with the champion effect, since
most players will stick to the subset of champions they are comfortable
with. The additional stability for the BA model comes from the champion
effect, which we can see by looking at the concordance index of the
champion effect, shown in Figure~\ref{fig-stab-champion}. In the EUW and
NA servers, concordance is typically above 0.8 for the champion effect.
Since the BA model does not distinguish between the champion effect and
the player effect, it will naturally be more stable.

We note that the player and champion effects are all relative
estimates---a value of 0 indicates the player (or champion) typically
performs at an average level, relative to all other players (or
champions). The low degree of intersection between the two data sets
then makes it more difficult to measure stability. The SIDO models still
have high stability for the player model regardless, but in the ally and
enemy models, where there are lower effect sizes and wider uncertainty
bounds, the models have relatively low stability. While the categories
help to improve stability, the improvement is modest compared to using
the model effects directly without the categorizations. However, the
ally/enemy models typically have higher stability in the champion
effect, as seen in Figure~\ref{fig-stab-champion}.

A full table of the independence scores is provided in the Appendix,
Section~\ref{sec-app-stab}.

\hypertarget{qualitative-analysis-and-examples}{%
\subsection{Qualitative analysis and
examples}\label{qualitative-analysis-and-examples}}

To evaluate the performance of our models, we assessed whether they
align with specific expectations pertaining to professional players.

\hypertarget{champion-prioritization}{%
\subsubsection{Champion prioritization}\label{champion-prioritization}}

Within a given season, certain champions in each role are deemed more
optimal for competitive play, leading players to dedicate increased time
to practicing and analyzing these champions during team practices and
competitive matches.

\begin{enumerate}
\def\labelenumi{\arabic{enumi}.}
\tightlist
\item
  \textbf{We examined the proficiency of professional players on the
  highest-priority champion for their respective roles during a season
  in comparison to the average non-professional player. We anticipated
  observing a significant difference in performance.}
\end{enumerate}

To conduct this assessment, we focused on professional bot-lane players
in the Korean server, as they are the most likely to practice the same
champions in solo queue as they would during team practices.

\begin{figure}[!ht]

{\centering \includegraphics{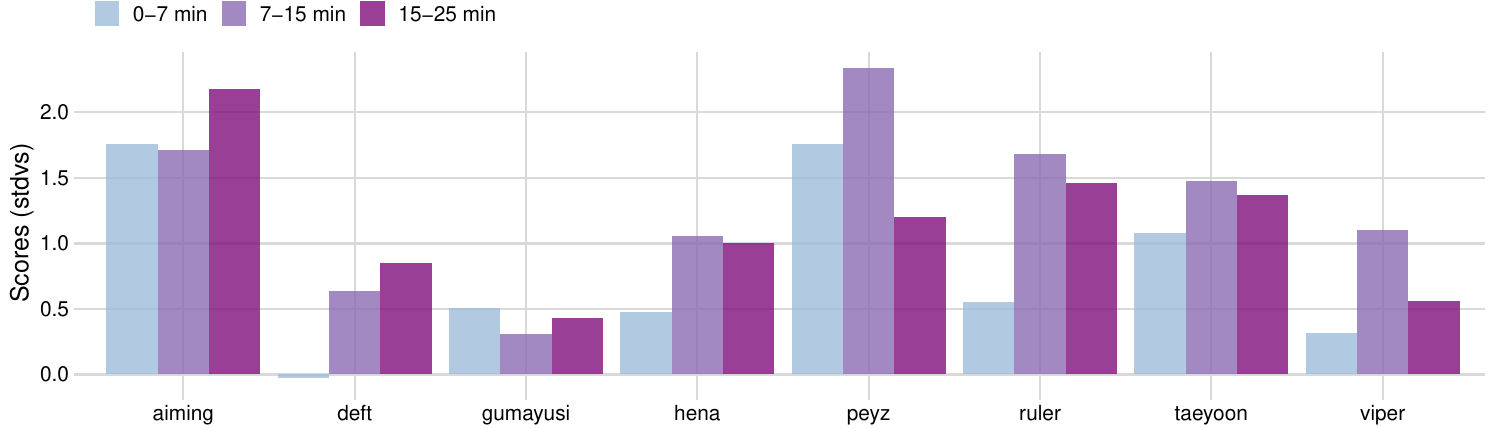}

}

\caption{\label{fig-kaisa-bot}LCK champion proficiencies (gold) on Kaisa
in the summer.}

\end{figure}

For instance, during the Summer 2023 season, Kaisa emerged as the most
prevalent bot lane pick (known as a ``meta'' pick) in competitive play
in Korea. We assumed that players extensively trained with this champion
during team practices. Figure~\ref{fig-kaisa-bot} illustrates that the
average proficiency of professional bot lane players on Kaisa, in terms
of both gold earned and damage dealt, was 1.075 and 1.1975,
respectively, as compared to non-professional accounts with an average
of 0.

\begin{enumerate}
\def\labelenumi{\arabic{enumi}.}
\setcounter{enumi}{1}
\tightlist
\item
  \textbf{We examined professional players' proficiency on champions
  they actively practiced in comparison to their proficiency on
  champions they did not prioritize. We anticipated an observable
  difference in performance.}
\end{enumerate}

\begin{figure}[!ht]

{\centering \includegraphics[width=0.85\textwidth,height=\textheight]{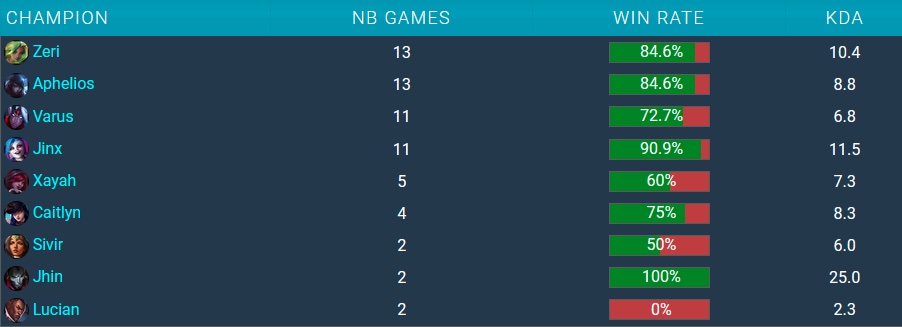}

}

\caption{\label{fig-ruler-golgg}Bot lane player Ruler's statistics in
Spring 2023. Screenshot from https://gol.gg.}

\end{figure}

To investigate this, we conducted an analysis focusing on Ruler,
consistently regarded as one of the best bot lane players globally, with
notable achievements including two regional championships and an MSI
title in 2023. During the Spring season, his preferred champions for
stage play encompassed Aphelios, Zeri, Varus, and Jinx, as shown in
Figure~\ref{fig-ruler-golgg}.

\begin{figure}[!ht]

{\centering \includegraphics{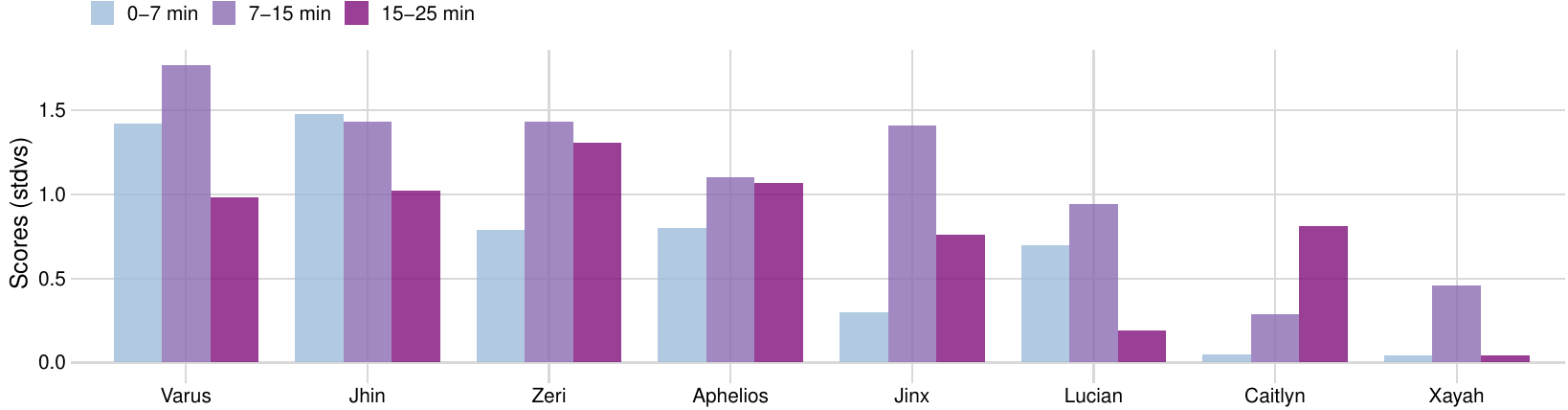}

}

\caption{\label{fig-ruler}Champion proficiency scores for Ruler in
spring.}

\end{figure}

Assuming that Ruler's on-stage champion selections reflect his practice
focus, we expected these champions to exhibit his highest proficiency in
solo queue. Figure~\ref{fig-ruler} shows that Ruler consistently
demonstrated significantly higher gold accumulation throughout all
phases of the game on champions he played on stage and presumably
practiced extensively, in contrast to those he did not prioritize.

\begin{enumerate}
\def\labelenumi{\arabic{enumi}.}
\setcounter{enumi}{2}
\tightlist
\item
  \textbf{We explored the proficiency levels of professional players on
  specific champions over time, considering the influence of increased
  or decreased practice.}
\end{enumerate}

To investigate this, we examined the proficiency change for Xayah among
professional bot lane players between the Spring and Summer seasons. We
selected this champion due to its increased presence, rising from 20\%
pick and ban rate in Spring to 59\% pick/ban rate in Summer, and the
availability of data for player performance on this champion in both
seasons. We expected an improvement in proficiency, assuming that
players had engaged in more consistent practice.

\begin{figure}[!ht]

{\centering \includegraphics{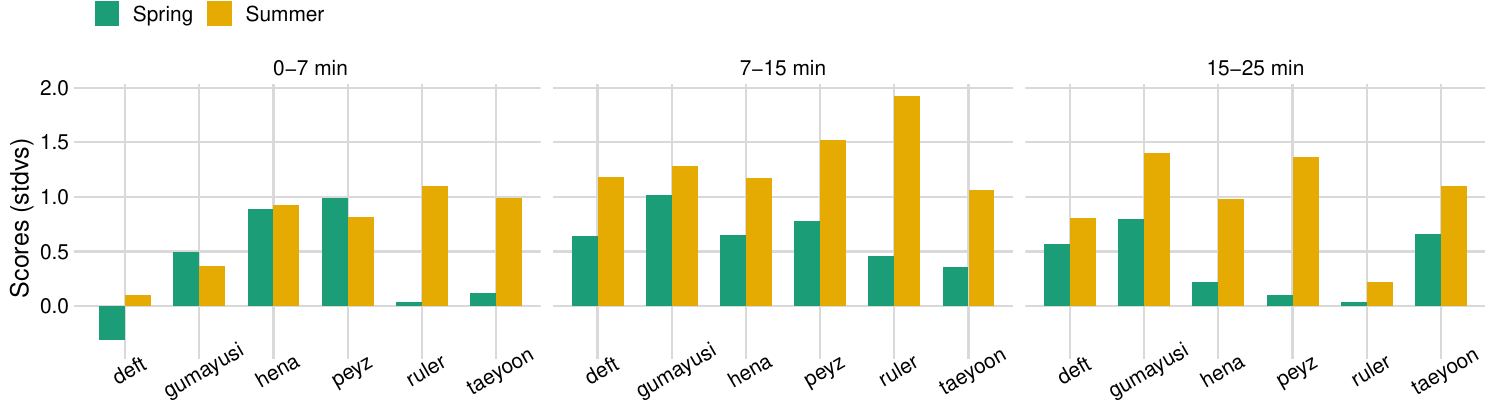}

}

\caption{\label{fig-xayah}Champion proficiency scores for LCK bot
players on Xayah. Aiming and V1per are not included due to insufficient
games on Xayah in spring.}

\end{figure}

As depicted in Figure~\ref{fig-xayah}, the average proficiency levels
for Xayah among each player demonstrated a notable increase from Spring
to Summer.

The three qualitative tests conducted in our study, while not
exhaustive, provide valuable insights into the model's capabilities with
regard to specific metrics. These tests underscore the model's ability
to:

\begin{itemize}
\item
  \textbf{Show player proficiency levels on different champions}: One of
  the primary strengths of our model is its capacity to reveal a
  player's proficiency on various champions and offer a more nuanced
  assessment of their skill set, transcending a one-size-fits-all
  evaluation.
\item
  \textbf{Compare different players}: The model's effectiveness in
  comparing different players is another noteworthy aspect. Through our
  analysis, we have demonstrated the model's ability to distinguish
  between professional players and non-professionals, showcasing its
  discriminative power. This capability opens avenues for evaluating
  player performance on a broader scale, allowing us to pinpoint
  standout individuals, identify trends, and make informed comparisons
  across the player spectrum.
\item
  \textbf{Tracking proficiency over time}: Our study highlights the
  model's effectiveness in tracking proficiency over time. This is vital
  for understanding player growth, adaptation, and development.
  Observing changes in proficiency offers insights into the impact of
  practice, evolving strategies, and meta shifts on a player's
  performance.
\end{itemize}

These findings lay the groundwork for our research, providing a robust
foundation for future endeavors. We aim to extend our models by
incorporating additional metrics that offer deeper insights into player
performance, such as adaptability and decision-making. Furthermore, our
aspiration is to broaden the scope of our research to encompass various
skill levels.

\hypertarget{collaborative-teamplay}{%
\subsubsection{Collaborative teamplay}\label{collaborative-teamplay}}

Another assumption we make is that professional players possess a better
strategic and tactical understanding of the game, that they are adept at
collaborating with teammates and responding effectively to opponents to
secure victory.

\begin{enumerate}
\def\labelenumi{\arabic{enumi}.}
\tightlist
\item
  \textbf{Analyzing Player Impact on Teammates and Enemies}:
\end{enumerate}

To assess this, we selected professional junglers from the Korean
server, as the role inherently affords players significant freedom
throughout the game in terms of where to go and what actions to
undertake. For instance, in the early game, a jungler can opt for the
most efficient gold-generating path, divert their efforts to aid a
struggling ally, or hinder and weaken an enemy player. Likewise, during
teamfights, the player can exercise discretion in choosing when to
maximize damage output or secure kills, determine the optimal timing for
enabling allies to inflict the most damage, or adopt strategic
positioning and attack patterns to thwart the most formidable foes.
These decisions, all equally valid, can pave the way for victory, with
the finest junglers adept at prioritizing them based on the game's
evolving state.

Our analysis encompassed various LCK junglers alongside Kanavi and
Tarzan, prominent figures from the Chinese league, and Blaber and
Jankos, esteemed junglers from North America and Europe respectively. We
anticipated a broad spectrum of playstyles and a discernible impact on
both allies and enemies compared to non-professional counterparts.

\begin{figure}[!ht]

{\centering \includegraphics{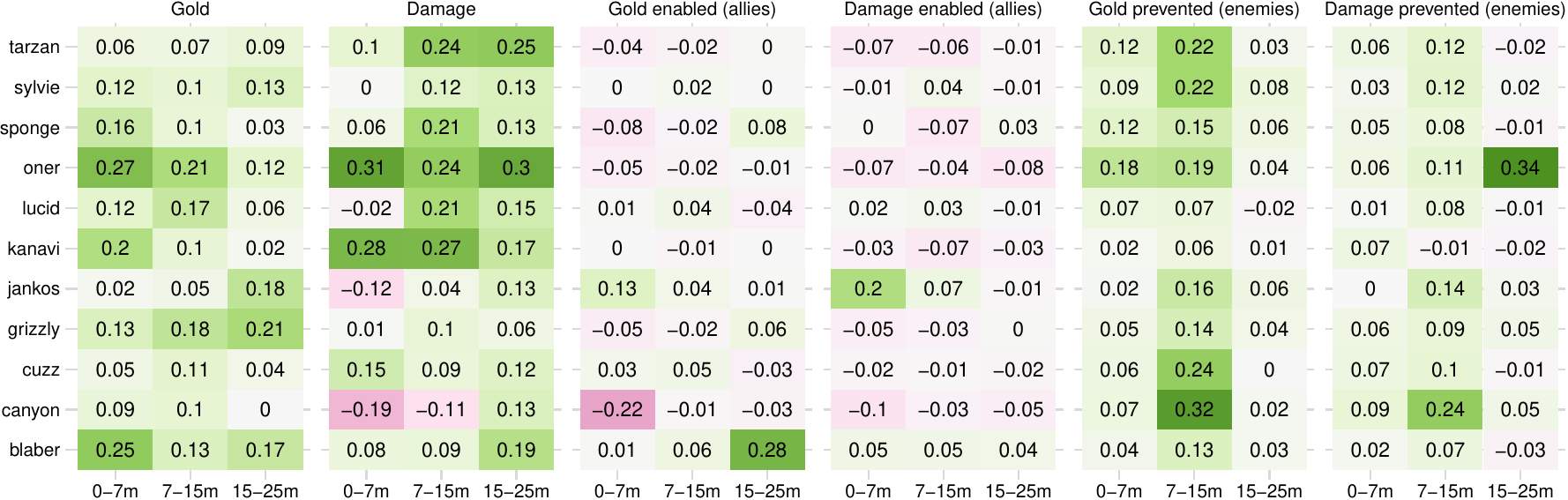}

}

\caption{\label{fig-stats-krjglrs}SIDO scores for LCK junglers
identified in our dataset, + Kanavi, Tarzan, Blaber, and Jankos. Scores
for players with multiple identified accounts are combined into one mean
score.}

\end{figure}

As depicted in Figure~\ref{fig-stats-krjglrs}, all professional junglers
exhibited proficiency in accumulating more gold. Notably, they also
excelled in impeding enemy gold acquisition and damage output in
comparison to their non-professional counterparts (with an average of
0). We posit that the impact of professional players on their allies may
be neutral or even negative when contrasted with their non-professional
counterparts. In the solo queue context, they might prioritize
self-resource allocation for victory, rather than diverting resources to
their non-professional allies.

\begin{enumerate}
\def\labelenumi{\arabic{enumi}.}
\setcounter{enumi}{1}
\tightlist
\item
  \textbf{Professional players distinguish themselves in terms of both
  technical skill and strategic/tactical depth.}
\end{enumerate}

To examine this distinction, we sought to quantify the extent of
difference in average proficiency levels between professional and
non-professional players across all measured categories. For the jungle
role, we anticipated that the most significant disparity between
professional and non-professional players would lie in their impact on
enemies rather than individual statistics. Conversely, for the bot lane
role, we expected the reverse trend due to the role's emphasis on
technical skill, damage capability, and limited strategic choices.

Initially, we conducted an analysis encompassing a broader spectrum of
professional and semi-professional junglers across the EU West and
Korean servers. Table~\ref{tbl-jglrs} illustrates that, on average, the
most substantial differentiation between professionals and
non-professionals lies in their impact on the enemy team's ability to
accumulate gold during the 7-15 minute phase.

\hypertarget{tbl-jglrs}{}
\begin{longtable}{lllllll}
\caption{\label{tbl-jglrs}Average SIDO model score for jungler accounts belonging to pros in the
LCK and LEC. Our database contains known accounts for 9 players in the
LCK and 7 players in the LEC. }\tabularnewline

\toprule
 & \multicolumn{3}{c}{LCK} & \multicolumn{3}{c}{LEC} \\ 
\cmidrule(lr){2-4} \cmidrule(lr){5-7}
Metric & 0-7m & 7-15m & 15-25m & 0-7m & 7-15m & 15-25m \\ 
\midrule\addlinespace[2.5pt]
Gold & 0.13 & 0.12 & 0.09 & 0.05 & 0.20 & 0.13 \\ 
Damage & 0.06 & 0.14 & 0.16 & 0.05 & 0.17 & 0.10 \\ 
Gold enabled (allies) & -0.03 & 0.01 & 0.03 & -0.08 & -0.03 & -0.02 \\ 
Damage enabled (allies) & -0.01 & -0.01 & -0.01 & -0.08 & -0.08 & -0.06 \\ 
Gold prevented (enemies) & 0.08 & 0.17 & 0.03 & 0.02 & 0.12 & 0.05 \\ 
Damage prevented (enemies) & 0.05 & 0.10 & 0.03 & 0.03 & 0.12 & 0.08 \\ 
\bottomrule
\end{longtable}

This variance can be attributed to the phase's complexity during the
laning phase, marked by numerous objectives to contest and dynamic
player movement across the map. Experienced professional junglers
demonstrate a keen ability to recognize diverse game states and make
well-informed decisions. However, this impact may diminish later in the
game when teams typically converge for major objectives, leaving fewer
choices for most roles overall. The jungle also has fewer resources than
the lanes and fewer opportunities to gain gold. This means the role
often has more limited capabilities in the 0-7 minute time frame and may
contribute to having less impact on allies and enemies in that period.

Furthermore, we extended our analysis to encompass all roles and all
players identified as belonging to a professional or semi-professional
team (including academy and amateur), then evaluated their average
proficiency levels across various categories. Building on our prior
analysis, we specifically focused on own gold earned and gold prevented
from enemies during the 0-7 minute and 7-15 minute intervals, as these
categories exhibited the most significant disparities between
professionals and non-professionals in the jungle role. Own gold earned
reflects more technical and mechanical skill, while gold prevented from
enemies reflects strategic and tactical decision-making. See
Section~\ref{sec-skill-categories} for examples.

\begin{figure}[!ht]

{\centering \includegraphics{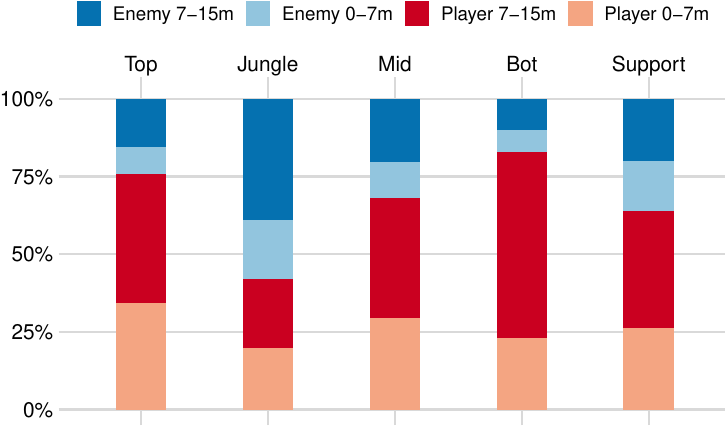}

}

\caption{\label{fig-decomp}Average pro and semi-pro player proficiency,
compared to non-pro proficiency, aggregated and presented as
percentages.}

\end{figure}

Figure~\ref{fig-decomp} highlights that, considering solely these two
categories, the discrepancy between technical skill and teamplay for bot
lane players leaned almost 2:1 in favor of technical skill. Conversely,
for junglers, it exhibited a 1:2 ratio, emphasizing the significance of
preventing enemies from accumulating gold. In contrast, for other roles,
the ratio was nearly 1:1.

\hypertarget{discussion}{%
\section{Discussion}\label{discussion}}

In this paper, we presented results comparing SIDO performance metrics
to a basic average model, representing the industry standard, and a
Plus-Minus model. We presented evidence that suggests the SIDO models
are better able to capture all the ways a player can contribute
positively to the game and not just to their own individual statistics,
something which the BA model cannot do. We showed as well that the
Plus-Minus model did not perform well in solo queue data, and posit that
the reason is because of the structure of solo queue data. In solo
queue, the same players may play multiple roles, which drastically
changes their statistics from game to game. Since the Plus-Minus model
does not differentiate between roles, this introduces additional noise.
The large number of players outside our data set that are required to be
included in the Plus-Minus model which may have skewed results.

We provided a number of qualitative examples showing how the SIDO models
can be used. We measured pro player proficiency on specific champions,
and we showed how this translated to competitive play. We demonstrated
how the SIDO models quantify strategic and tactical skill by comparing
professional junglers to an average grandmaster player, and showed the
majority of junglers performed well at shutting down their enemies, in
addition to performing well in terms of their own statistics. Lastly, we
showed that the jungle and support roles demonstrate more strategic and
tactical skill as compared to the other roles. Notably, the bot role has
the heaviest emphasis on mechanics and technical skill.

Although the SIDO ally and enemy models in many ways appear to perform
as desired---they are able to capture differences between professional
players at the highest level of competitive play relative to average
high-MMR soloqueue players and provide context to player statistics that
conforms with expectations---low discrimination scores suggest taking
any numerical result with a note of caution. In general, player effect
sizes for the ally/enemy models are lower than for the player models,
which naturally leads to lower discrimination. This may be a result of
soloqueue play, which does not have the same level of communication as
teams in competitive play and so cooperation within a team is more
difficult. Future work is needed to investigate this further.

The SIDO models represent first steps towards framing player performance
as the combination of their total impact on the game (themselves, their
allies, and their enemies). In this way, we hope to shift the discourse
towards a more holistic understanding of the game and enable:

\begin{enumerate}
\def\labelenumi{\arabic{enumi}.}
\tightlist
\item
  \textbf{Better experience for players}
\end{enumerate}

In the world of professional gaming, solo queue is where most players
sharpen their skills. But solo queue isn't a perfect reflection of
teamwork, as players can't plan or talk to their teammates as
effectively as in team practices or real competitions. So, players often
think that to win in solo queue, they have to do everything themselves.
When they fall behind and can't carry the game, they take big risks to
try and get back in control. When many players share this mindset, it
can lead to conflicts within the team. This focus on individual
performance is also encouraged by the data shown to players and
analysts. They often look at individual stats to compare players, and
the post-game screen in League of Legends emphasizes individual
achievements.

But our research shows that professional players stand out not just for
their individual stats but also for how they influence their
teammates---a kind of teamwork that's possible even in solo queue.

To bridge this gap, we need to introduce new measures that highlight how
a player affects their teammates and opponents. This change could
encourage players to value teamwork more and make more strategic and
cooperative decisions in the game. This shift towards teamwork aligns
with what we see in professional play and could make the gaming
experience more enjoyable and strategic for everyone.

\begin{enumerate}
\def\labelenumi{\arabic{enumi}.}
\setcounter{enumi}{1}
\tightlist
\item
  \textbf{Scouting at the professional level}
\end{enumerate}

The issue of superteams in professional gaming is a recurring phenomenon
during each offseason. Teams invest significant resources in creating
these superteams by assembling players with exceptional technical skills
from various professional teams. However, there is a fundamental problem
with this approach. When these individual star players were part of
their previous teams, they often enjoyed certain privileges and the
freedom to play the game in a way that best suited their strengths.
Their former teammates understood how to enable them, and this synergy
allowed them to shine.

The real challenge arises when these star players are brought together
into a single team. Despite their remarkable individual skills, they may
have distinct playing styles that do not necessarily align. Moreover,
they lack the experience of enabling their new teammates in the same way
they were supported in their previous teams. This can lead to a
disconnect within the superteam, resulting in underperformance compared
to what would be expected based on the individual talents of its
members.

A more effective approach for organizations could involve considering
players as a combination of their technical skill and teamplay. By
evaluating a player's ability to work cohesively with their teammates
and contribute to the overall team strategy, organizations can build
more harmonious and efficient teams without the need for exorbitant
budgets. This approach not only ensures better team dynamics but also
maximizes the potential of each player within the collective unit,
ultimately leading to more successful and balanced teams in the
competitive gaming landscape. In this way, team synergy and teamwork
become critical factors in creating a winning formula, rather than just
relying on individual star power.

\begin{enumerate}
\def\labelenumi{\arabic{enumi}.}
\setcounter{enumi}{2}
\tightlist
\item
  \textbf{More efficient coaching at all levels of the ecosystem}
\end{enumerate}

A significant challenge within the League of Legends ecosystem lies in
the relatively primitive state of coaching, particularly with a heavy
focus on performance rather than player development. One of the primary
reasons behind this limitation is the lack of objective tools to track
player performances and proficiencies in a quantifiable manner. In the
absence of such metrics, coaches often rely on subjective evaluations
and anecdotal evidence to assess players' strengths and weaknesses.

This issue is particularly pronounced in a game like League of Legends,
which is still relatively young in the realm of competitive esports.
Unlike more established sports with well-established coaching
methodologies and data-driven approaches, League of Legends coaching has
lagged behind in terms of sophistication. The current emphasis on
performance rather than player development can hinder the long-term
growth and potential of players. The solution to this challenge lies in
creating a quantifiable way to evaluate players in different roles at
all skill levels. By developing comprehensive metrics and data-driven
assessments that go beyond individual statistics, coaches can gain a
deeper understanding of a player's proficiency, decision-making, and
overall impact on the game. These metrics can encompass not only
technical skills but also teamplay, adaptability, and strategic
thinking.

With such tools at their disposal, coaches can design more efficient and
tailored training regimens to help their players learn, improve, and
perform consistently. By identifying areas for improvement and tracking
progress over time, coaches can provide targeted guidance and support,
fostering player development and elevating the overall competitive
landscape. This shift towards a more data-driven and objective approach
to coaching holds the potential to revolutionize player development in
League of Legends and contribute to the game's continued growth and
evolution as a competitive esport.

\begin{enumerate}
\def\labelenumi{\arabic{enumi}.}
\setcounter{enumi}{3}
\tightlist
\item
  \textbf{A metric that allows for research projects to be conducted at
  scale}
\end{enumerate}

Conducting research in competitive gaming has long been hindered by the
absence of a standardized metric to evaluate player performance. This
challenge has raised questions about how various factors, such as sleep
patterns or practice routines, impact player performance. Traditional
research methods often involve subjective assessments, resulting in
potential biases and limited generalizability.

A model that benchmarks player's proficiency and performance in solo
queue that can be tracked over a period of time can allow for:

\begin{itemize}
\item
  Objective analysis: Researchers can now employ a more objective
  framework for assessing player performance.
\item
  Precise variable measurement: The model enables more precise
  measurements of different performance factors, ensuring the accuracy
  of research findings.
\item
  Comparative research: Researchers can conduct direct comparisons
  across players, teams, or regions, uncovering valuable insights into
  performance disparities and trends.
\item
  Longitudinal studies: The model facilitates longitudinal studies,
  allowing researchers to explore the developmental trajectories of
  player skills and strategies over time.
\item
  Causal inference: Researchers can investigate causal relationships,
  providing a more robust basis for drawing conclusions.
\item
  Sample size and generalizability: The model's scalability accommodates
  large datasets, increasing the statistical power and generalizability
  of research.
\item
  Hypothesis testing: Researchers can rigorously test hypotheses
  regarding the impact of various factors on player performance.
\end{itemize}

\hypertarget{future-work}{%
\section{Future work}\label{future-work}}

Our exploratory journey into the realm of performance modeling in
competitive gaming has illuminated numerous exciting avenues for future
research and development. While our current work serves as a
foundational stepping stone, there exists a vast landscape of
possibilities waiting to be explored. Here, we outline three primary
directions in which we intend to expand, validate, and apply our
performance models in the coming year.

\begin{itemize}
\item
  \textbf{Expansion}: The first path forward involves a substantial
  expansion of our models. To create a more comprehensive and inclusive
  framework, we aim to encompass a broader spectrum of the playerbase.
  This entails incorporating a wider array of metrics that capture
  additional dimensions of player performance. Notably, we recognize the
  significance of vision as an independent and vital aspect of the game,
  and its incorporation will enhance our model's representation of the
  support role. By broadening our horizons, we aspire to provide a more
  holistic understanding of player proficiency.
\item
  \textbf{Validation}: A critical aspect of our future work centers
  around refining the validation process. We acknowledge the limitations
  of our current assumptions regarding player practice regimens,
  priorities, and perceptions of solo queue. To bridge the gap between
  solo queue and competitive play more effectively, we plan to establish
  partnerships with professional and semi-professional teams. This
  collaboration will enable us to gather invaluable data from team
  practices, offering a richer and more authentic perspective. In
  particular, practice games will provide a larger sample size and more
  granular data, enabling us to refine our assumptions. Insights gained
  from examining the interplay between professional players' solo queue,
  team practices, and on-stage performances will inform the development
  of programs at the amateur, collegiate, and community levels.
\item
  \textbf{Application}: The ultimate goal of our research is to
  translate our models into actionable insights and practical
  applications. We envision conducting experiments that investigate the
  impact of crucial factors, such as sleep and communication, on the
  learning, training, and competitive performances of players. Sleep, a
  fundamental health factor, profoundly affects cognitive load capacity.
  Surprisingly, even professional teams have yet to incorporate sleep
  tracking into their programs. Similarly, efficient in-game
  communication, a crucial aspect of competitive gaming, lacks a
  standardized approach. Our aim is to establish a direct relationship
  between these variables and players' learning speed and performance.
  Leveraging our performance metrics, we intend to compare a larger
  player cohort to identify key patterns and correlations.
\end{itemize}

In summary, our future work is poised to expand the horizons of
performance modeling in competitive gaming. By embracing inclusivity,
refining validation methods, and unlocking practical applications, we
endeavor to propel the field forward. Our journey is not only one of
discovery but also of innovation, with the ultimate aspiration of
elevating the standard of research, enhancing player development, and
fostering a more informed and efficient competitive gaming ecosystem.

\hypertarget{references}{%
\section{References}\label{references}}

\hypertarget{refs}{}
\begin{CSLReferences}{1}{0}
\leavevmode\vadjust pre{\hypertarget{ref-aung2018}{}}%
Aung, M., V. Bonometti, A. Drachen, P. Cowling, A. V. Kokkinakis, C.
Yoder, and A. Wade. 2018. {``Predicting Skill Learning in a Large,
Longitudinal MOBA Dataset.''} In \emph{2018 IEEE Conference on
Computational Intelligence and Games (CIG)}, 1--7. Maastricht: IEEE.
\url{https://doi.org/10.1109/CIG.2018.8490431}.

\leavevmode\vadjust pre{\hypertarget{ref-bahrololloomi2022}{}}%
Bahrololloomi, Farnod, Sebastian Sauer, Fabio Klonowski, Robin Horst,
and Ralf Dörner. 2022. {``A Machine Learning Based Analysis of e-Sports
Player Performances in League of Legends for Winning Prediction Based on
Player Roles and Performances.''} In \emph{6th International Conference
on Human Computer Interaction Theory and Applications}, 68--76.
SCITEPRESS - Science; Technology Publications.
\url{https://doi.org/10.5220/0010895900003124}.

\leavevmode\vadjust pre{\hypertarget{ref-baker2018}{}}%
Baker, Thomas. 2018. {``Standard League of Legends Statistics
Oversimplify the Game.''} \emph{Game Haus}, March.
\url{https://thegamehaus.com/columns/standard-league-of-legends-statistics-oversimplify-the-game/2018/03/15/}.

\leavevmode\vadjust pre{\hypertarget{ref-benjamini1995}{}}%
Benjamini, Yoav, and Yosef Hochberg. 1995. {``Controlling the False
Discovery Rate: A Practical and Powerful Approach to Multiple
Testing.''} \emph{Journal of the Royal Statistical Society: Series B
(Methodological)} 57 (1): 289--300.
\url{https://doi.org/10.1111/j.2517-6161.1995.tb02031.x}.

\leavevmode\vadjust pre{\hypertarget{ref-birant2022}{}}%
Birant, Kokten Ulas, and Derya Birant. 2022. {``Multi-Objective
Multi-Instance Learning: A New Approach to Machine Learning for
eSports.''} \emph{Entropy} 25 (1): 28.
\url{https://doi.org/10.3390/e25010028}.

\leavevmode\vadjust pre{\hypertarget{ref-clark2020}{}}%
Clark, Nicholas, Brian Macdonald, and Ian Kloo. 2020. {``A Bayesian
Adjusted Plus-Minus Analysis for the Esport Dota 2.''} \emph{Journal of
Quantitative Analysis in Sports} 16 (4): 325--41.
\url{https://doi.org/10.1515/jqas-2019-0103}.

\leavevmode\vadjust pre{\hypertarget{ref-dehpanah2021}{}}%
Dehpanah, Arman, Muheeb Faizan Ghori, Jonathan Gemmell, and Bamshad
Mobasher. 2021. {``Evaluating Team Skill Aggregation in Online
Competitive Games.''} In \emph{2021 IEEE Conference on Games (CoG)},
01--08. Copenhagen, Denmark: IEEE.
\url{https://doi.org/10.1109/CoG52621.2021.9618994}.

\leavevmode\vadjust pre{\hypertarget{ref-do2021}{}}%
Do, Tiffany D., Seong Ioi Wang, Dylan S. Yu, Matthew G. McMillian, and
Ryan P. McMahan. 2021. {``Using Machine Learning to Predict Game
Outcomes Based on Player-Champion Experience in League of Legends.''} In
\emph{FDG'21: The 16th International Conference on the Foundations of
Digital Games 2021}, 1--5. Montreal QC Canada: ACM.
\url{https://doi.org/10.1145/3472538.3472579}.

\leavevmode\vadjust pre{\hypertarget{ref-drachen2014}{}}%
Drachen, Anders, Matthew Yancey, John Maguire, Derrek Chu, Iris Yuhui
Wang, Tobias Mahlmann, Matthias Schubert, and Diego Klabajan. 2014.
{``Skill-Based Differences in Spatio-Temporal Team Behaviour in Defence
of the Ancients 2 (DotA 2).''} In \emph{2014 IEEE Games, Media,
Entertainment (GEM) Conference}, 1--8. Toronto, ON: IEEE.
\url{https://doi.org/10.1109/GEM.2014.7048109}.

\leavevmode\vadjust pre{\hypertarget{ref-elo1967}{}}%
Elo, Arpad E. 1967. {``The Proposed USCF Rating System: Its Development,
Theory and Applications.''} \emph{Chess Life} 22 (August): 242--47.
\url{https://uscf1-nyc1.aodhosting.com/CL-AND-CR-ALL/CL-ALL/1967/1967_08.pdf\#page=26}.

\leavevmode\vadjust pre{\hypertarget{ref-franks2016}{}}%
Franks, Alexander M., Alexander D'Amour, Daniel Cervone, and Luke Bornn.
2016. {``Meta-Analytics: Tools for Understanding the Statistical
Properties of Sports Metrics.''} \emph{Journal of Quantitative Analysis
in Sports} 12 (4). \url{https://doi.org/10.1515/jqas-2016-0098}.

\leavevmode\vadjust pre{\hypertarget{ref-gelman2006}{}}%
Gelman, Andrew. 2006. {``Prior Distributions for Variance Parameters in
Hierarchical Models (Comment on Article by Browne and Draper).''}
\emph{Bayesian Analysis} 1 (3). \url{https://doi.org/10.1214/06-BA117A}.

\leavevmode\vadjust pre{\hypertarget{ref-gelman2017}{}}%
Gelman, Andrew, Daniel Simpson, and Michael Betancourt. 2017. {``The
Prior Can Often Only Be Understood in the Context of the Likelihood.''}
\emph{Entropy} 19 (10): 555. \url{https://doi.org/10.3390/e19100555}.

\leavevmode\vadjust pre{\hypertarget{ref-gerrard2007}{}}%
Gerrard, Bill. 2007. {``Is the Moneyball Approach Transferable to
Complex Invasion Team Sports?''} \emph{International Journal of Sport
Finance} 2 (4): 214--30.
\url{https://www.proquest.com/docview/229399553?pq-origsite=gscholar\&fromopenview=true}.

\leavevmode\vadjust pre{\hypertarget{ref-hajir2021}{}}%
Hajir, Alexander. 2021. {``Significant Statistics: A Basic Guide to
Informing Your Opinions with Oracle{'}s Elixir.''}
\url{https://oracleselixir.com/blog/post/13483/significant-statistics-a-basic-guide-to-informing-your-opinions-with-oracles-elixir}.

\leavevmode\vadjust pre{\hypertarget{ref-herbrich2007}{}}%
Herbrich, Ralf, Tom Minka, and Thore Graepel. 2007.
{``TrueSkill{\texttrademark}: A Bayesian Skill Rating System.''} In,
edited by Bernhard Schölkopf, John Platt, and Thomas Hofmann, 569--76.
The MIT Press. \url{https://doi.org/10.7551/mitpress/7503.003.0076}.

\leavevmode\vadjust pre{\hypertarget{ref-hvattum2019}{}}%
Hvattum, Lars Magnus. 2019. {``A Comprehensive Review of Plus-Minus
Ratings for Evaluating Individual Players in Team Sports.''}
\emph{International Journal of Computer Science in Sport} 18: 1--23.
\url{https://doi.org/10.2478/ijcss-2019-0001}.

\leavevmode\vadjust pre{\hypertarget{ref-jeong2022}{}}%
Jeong, Inhyeok, Kento Nakagawa, Rieko Osu, and Kazuyuki Kanosue. 2022.
{``Difference in Gaze Control Ability Between Low and High Skill Players
of a Real-Time Strategy Game in Esports.''} Edited by Greg Wood.
\emph{PLOS ONE} 17 (3): e0265526.
\url{https://doi.org/10.1371/journal.pone.0265526}.

\leavevmode\vadjust pre{\hypertarget{ref-kahn2016}{}}%
Kahn, Adam S., and Dmitri Williams. 2016. {``We{'}re All in This (Game)
Together: Transactive Memory Systems, Social Presence, and Team
Structure in Multiplayer Online Battle Arenas.''} \emph{Communication
Research} 43 (4): 487--517.
\url{https://doi.org/10.1177/0093650215617504}.

\leavevmode\vadjust pre{\hypertarget{ref-kaplan2014}{}}%
Kaplan, Edward H., Kevin Mongeon, and John T. Ryan. 2014. {``A Markov
Model for Hockey: Manpower Differential and Win Probability Added.''}
\emph{INFOR: Information Systems and Operational Research} 52 (2):
39--50. \url{https://doi.org/10.3138/infor.52.2.39}.

\leavevmode\vadjust pre{\hypertarget{ref-kim2020}{}}%
Kim, Dong-Hee, Changwoo Lee, and Ki-Seok Chung. 2020. {``A
Confidence-Calibrated MOBA Game Winner Predictor.''} In, 622625. Osaka,
Japan: IEEE. \url{https://doi.org/10.1109/CoG47356.2020.9231878}.

\leavevmode\vadjust pre{\hypertarget{ref-kim2017}{}}%
Kim, Young Ji, David Engel, Anita Williams Woolley, Jeffrey Yu-Ting Lin,
Naomi McArthur, and Thomas W. Malone. 2017. {``What Makes a Strong
Team?: Using Collective Intelligence to Predict Team Performance in
League of Legends.''} In \emph{CSCW '17: Computer Supported Cooperative
Work and Social Computing}, 2316--29. Portland Oregon USA: ACM.
\url{https://doi.org/10.1145/2998181.2998185}.

\leavevmode\vadjust pre{\hypertarget{ref-lewis2004}{}}%
Lewis, Michael. 2004. \emph{Moneyball: The Art of Winning an Unfair
Game}. WW Norton \& Company.

\leavevmode\vadjust pre{\hypertarget{ref-liang2021}{}}%
Liang, Muchen. 2021. {``Research on Prediction of the Game Winner Based
on Artificial Intelligence Methods.''} In \emph{ICAIP 2021: 2021 5th
International Conference on Advances in Image Processing}, 97--102.
Chengdu China: ACM. \url{https://doi.org/10.1145/3502827.3502843}.

\leavevmode\vadjust pre{\hypertarget{ref-lolesports2017}{}}%
LoL Esports. 2017. {``Stats Science 101: EU LCS Top Laner
Performance.''} \url{https://www.youtube.com/watch?v=kkp_LMYBEFY}.

\leavevmode\vadjust pre{\hypertarget{ref-mengersen2016}{}}%
Mengersen, Kerrie L., Christopher C. Drovandi, Christian P. Robert,
David B. Pyne, and Christopher J. Gore. 2016. {``Bayesian Estimation of
Small Effects in Exercise and Sports Science.''} Edited by Cathy W. S.
Chen. \emph{PLOS ONE} 11 (4): e0147311.
\url{https://doi.org/10.1371/journal.pone.0147311}.

\leavevmode\vadjust pre{\hypertarget{ref-missfortunedabes2022}{}}%
MissFortuneDaBes. 2022. {``League of Legends Stat Websites Are LYING to
You. Here's How.''} \url{https://www.youtube.com/watch?v=6pzrI3WqQTY}.

\leavevmode\vadjust pre{\hypertarget{ref-morris1983}{}}%
Morris, Carl N. 1983. {``Parametric Empirical Bayes Inference: Theory
and Applications.''} \emph{Journal of the American Statistical
Association} 78 (381): 47--55.
\url{https://doi.org/10.1080/01621459.1983.10477920}.

\leavevmode\vadjust pre{\hypertarget{ref-pettigrew2015}{}}%
Pettigrew, Stephen. 2015. {``Assessing the Offensive Productivity of NHL
Players Using in-Game Win Probabilities.''} In, 2:8.

\leavevmode\vadjust pre{\hypertarget{ref-polson2012}{}}%
Polson, Nicholas G., and James G. Scott. 2012. {``On the Half-Cauchy
Prior for a Global Scale Parameter.''} \emph{Bayesian Analysis} 7 (4).
\url{https://doi.org/10.1214/12-BA730}.

\leavevmode\vadjust pre{\hypertarget{ref-santos-fernandez2019}{}}%
Santos-Fernandez, Edgar, Paul Wu, and Kerrie L. Mengersen. 2019.
{``Bayesian Statistics Meets Sports: A Comprehensive Review.''}
\emph{Journal of Quantitative Analysis in Sports} 15 (4): 289--312.
\url{https://doi.org/10.1515/jqas-2018-0106}.

\leavevmode\vadjust pre{\hypertarget{ref-sapienza2019}{}}%
Sapienza, Anna, Palash Goyal, and Emilio Ferrara. 2019. {``Deep Neural
Networks for Optimal Team Composition.''} \emph{Frontiers in Big Data} 2
(June): 14. \url{https://doi.org/10.3389/fdata.2019.00014}.

\leavevmode\vadjust pre{\hypertarget{ref-suznjevic2015}{}}%
Suznjevic, Mirko, Maja Matijasevic, and Jelena Konfic. 2015.
{``Application Context Based Algorithm for Player Skill Evaluation in
MOBA Games.''} In \emph{2015 International Workshop on Network and
Systems Support for Games (NetGames)}, 1--6. Zagreb, Croatia: IEEE.
\url{https://doi.org/10.1109/NetGames.2015.7382993}.

\leavevmode\vadjust pre{\hypertarget{ref-toth2021}{}}%
Toth, Adam J., Niall Ramsbottom, Christophe Constantin, Alain Milliet,
and Mark J. Campbell. 2021. {``The Effect of Expertise, Training and
Neurostimulation on Sensory-Motor Skill in Esports.''} \emph{Computers
in Human Behavior} 121: 106782.
\url{https://doi.org/10.1016/j.chb.2021.106782}.

\leavevmode\vadjust pre{\hypertarget{ref-velichkovsky2019}{}}%
Velichkovsky, Boris B., Nikita Khromov, Alexander Korotin, Evgeny
Burnaev, and Andrey Somov. 2019. {``Visual Fixations Duration as an
Indicator of Skill Level in eSports.''} In, edited by David Lamas,
Fernando Loizides, Lennart Nacke, Helen Petrie, Marco Winckler, and
Panayiotis Zaphiris, 11746:397--405. Cham: Springer International
Publishing.
\url{https://link.springer.com/10.1007/978-3-030-29381-9_25}.

\leavevmode\vadjust pre{\hypertarget{ref-xia2019}{}}%
Xia, Bang, Huiwen Wang, and Ronggang Zhou. 2019. {``What Contributes to
Success in MOBA Games? An Empirical Study of Defense of the Ancients
2.''} \emph{Games and Culture} 14 (5): 498--522.
\url{https://doi.org/10.1177/1555412017710599}.

\leavevmode\vadjust pre{\hypertarget{ref-zhang2023}{}}%
Zhang, Amy X., Le Bao, Changcheng Li, and Michael Daniels. 2023.
{``Approximate Cross-Validated Mean Estimates for Bayesian Hierarchical
Regression Models.''} \emph{arXiv}.
\url{https://arxiv.org/abs/2011.14238}.

\leavevmode\vadjust pre{\hypertarget{ref-zhang2017}{}}%
Zhang, Xiaoling, Yufeng Yue, Xiaofei Gu, Ben Niu, and Y. Y. Feng. 2017.
{``Investigating the Impact of Champion Features and Player Information
on Champion Usage in League of Legends.''} In \emph{ICIT 2017: 2017
International Conference on Information Technology}, 91--95. Singapore
Singapore: ACM. \url{https://doi.org/10.1145/3176653.3176730}.

\end{CSLReferences}

\hypertarget{appendix}{%
\section{Appendix}\label{appendix}}

\hypertarget{sec-glossary}{%
\subsection{League of Legends Glossary}\label{sec-glossary}}

\textbf{Armor:} An in-game statistic for the player's character that
contributes to reduction of incoming physical-type damage. See also
Magic Resist.

\textbf{Assist:} Contributing to the killing blow of an enemy champion.
Assists occur if one player helps deal damage and/or applies crowd
control to the enemy champion, or heals/buffs their ally in a short
window before the killing blow.

\textbf{Baron Nashor (Baron):} A neutral monster in the top-side jungle
of the map. Defeating Baron grants a temporary buff and is often a
pivotal objective in the late game.

\textbf{Bot Lane:} See Roles: Bot.

\textbf{Buffs:} Temporary enhancements gained from various sources,
boosting specific stats or granting special abilities.

\textbf{Carries:} Champions whose primary role is to inflict high damage
on enemy champions and objectives.

\textbf{Challenger (rank):} The name for the top-ranked players on each
server, based on LP. See also LP.

\textbf{CS} \textbf{(Creep Score):} The number of last hits landed on
enemy minions (small-scale monsters which travel through the lanes and
help defend towers), which provides gold and experience.

\textbf{DotA:} Defense of the Ancients 2, another popular MOBA often
compared to League of Legends.

\textbf{Drake (Dragon)}: A neutral monster in the bottom-side jungle of
the map, which grants temporary buffs. Typically contested throughout
the course of the game.

\textbf{Early Game:} The first phase of the game, where the top, middle,
bot, and support players face off against their lane opponents at the
boundaries of their territory, between two sets of opposing towers. The
only significant variation at this stage comes from the jungle role, the
only role which does not go to a tower at the beginning of the game and
instead roams in the jungle area between lanes.

\textbf{Elo:} A rating system for calculating the relative skill level
of players in competitive games, typically zero-sum games like chess,
go, and many competitive video games.

\textbf{Grandmaster (rank):} The second-highest tier for players on each
server, after challenger. In NA, KR, and EUW, a player must be in the
top 1000 players, based on LP, to make it to grandmaster tier. See also
challenger, LP.

\textbf{Herald (Rift Herald):} A neutral monster in the top-side jungle
of the map. Defeating Herald grants a buff and control of a temporary
allied Herald which can aid in taking down enemy turrets in the
mid-game.

\textbf{Items}: Equipment a player can purchase within the game using
gold earned from various sources. Items provide stat boosts, unique
enhancements, and active abilities, allowing you to customize your
champion's strengths and adapt to different situations.

\textbf{Jungle:} The neutral area between lanes where a champion roams,
clears monsters, and ganks lanes. May also refer to the player role, see
Roles: Jungler.

\textbf{Kill}: Dealing the final blow to an enemy champion.

\textbf{LCS:} League of Legends Championship Series, the top-level North
American professional league.

\textbf{LEC:} League of Legends European Championship, the top-level
European professional league.

\textbf{LCK:} League of Legends Champions Korea, the top-level South
Korean professional league.

\textbf{LPL:} League of Legends Pro League, the top-level Chinese
professional league.

\textbf{Lane}: One out of three designated paths (top, middle, bot) on
the map where champions primarily duel or farm against their opponent.
Lanes provide structure and focus for early game competition, offering
experience, gold, and objectives like towers to fight over.

\textbf{Lane Advantage:} Having superior pressure, resources, or control
over the opponent in your lane.

\textbf{Laning Phase:} The early phase where champions compete for
resources and experience in their assigned lanes. See Early Game.

\textbf{Late Game:} The final phase of the game, characterized by more
extensive teamfights focused around major objectives. By this time
point, the laning phase has concluded and all tower plates have been
removed.

\textbf{LP (League Points):} In solo queue, a number ranging from 0 to
100 that represents progress towards a higher ranked tier. In the
master, grandmaster, and challenger tiers, the value of LP is unbounded
and is used to determine exact rankings. The top 1000 players by LP are
split into the grandmaster and challenger tiers. Winning a match causes
a gain in LP, while losing a match causes a reduction in LP. LP is
related to MMR, but unlike MMR, it is not hidden.

\textbf{Magic Resist:} An in-game statistic for the player's character
that contributes to reduction of incoming magical-type damage. See also
Armor.

\textbf{Map:} The entire playing field in a MOBA. In this paper, ``map''
refers specifically to Summoner's Rift, though other game modes have
different maps.

\textbf{Map Control:} Controlling key areas on the map to limit enemy
movement and access to resources.

\textbf{Map Impact:} A champion's ability to influence the entire map
through pressure, mobility, or global abilities.

\textbf{Mid Game:} The transitional phase between the laning phase of
the early game and the open map of the late game, where objectives
become more crucial and teamfights become more common.

\textbf{MMR:} Matchmaking Rating, the hidden rating used to match
players of similar skill level.

\textbf{MOBA:} Multiplayer Online Battle Arena, a genre of video game
that pits two teams of players against each other in an online
battlefield.

\textbf{Neutral Objective:} Monsters and areas outside lanes that offer
strategic advantages, like buffs or vision, and do not belong to any
team.

\textbf{Nexus:} Each team's base structure, located on opposite corners
of the map. Destroying the enemy team's Nexus wins the game.

\textbf{Objective:} Any point of interest on the map worth contesting,
like towers, buffs, dragons, or Rift Herald.

\textbf{Pathing:} The route a jungler takes through the jungle,
optimizing monster clears and gank opportunities.

\textbf{Pick/Ban}: A phase at the beginning of the match where teams
take turns banning champions they don't want their opponents to play and
selecting their own champions. Picking strategically for team synergy
and countering enemy picks is crucial for setting the stage for the
game.

\textbf{Ranked Ladder:} The leaderboard displaying players' ranks and
their relative position within their region.

\textbf{Rift Herald:} A neutral monster granting a powerful buff to a
tower upon its destruction.

\textbf{Roam:} Leaving your lane to assist teammates in other lanes or
secure objectives.

\textbf{Roles:} Similar to positions within traditional sports, players
elect to play one of five roles which correspond both to a high-level
playstyle and a zone of the battlefield the player is responsible for in
the early game.

\begin{itemize}
\item
  \textbf{Top:} Typically plays self-sufficient champions and starts off
  the game in the top-most lane.
\item
  \textbf{Jungler:} Travels in the jungle area between lanes. The
  jungler's actions are often not visible to the enemy team and the
  jungler's role is often to coordinate with their teammates in planning
  ambushes.
\item
  \textbf{Mid:} Typically plays a mage-type champion and starts off the
  game in the middle lane. The mid role may also roam to the side lanes.
\item
  \textbf{Bot:} Typically plays fragile damage dealers and starts off
  the game in the bottom-most lane, aided by the support player.
\item
  \textbf{Support:} Typically plays utility-focused champions with the
  goal of aiding mainly their bot lane partner in the early game and
  team in general.
\end{itemize}

\textbf{Skillshot:} An ability that requires aim and skill to hit a
target, as opposed to abilities which automatically hit their targets.

\textbf{Solo Queue:} A play mode where players are individually matched
based on their skill level, using the hidden internal matchmaking rating
(see MMR). Match results contribute to a player's rank on the ranked
ladder.

\textbf{Support:} See Roles: Support.

\textbf{Summoner's Rift:} The game's primary map, a mirrored battlefield
divided by jungle areas and three lanes.

\textbf{Tanks:} Champions with high health and defenses, designed to
absorb damage and disrupt enemy teams.

\textbf{Team Composition:} The mix of champions a team chooses, focusing
on synergy and overall strengths.

\textbf{Teamfights:} Large-scale clashes between teams, often
determining the momentum of the game.

\textbf{Top Lane:} See Roles: Top.

\textbf{Towers:} Defensive structures guarding each lane, offering
vision and gold/experience upon destruction.

\textbf{Tower Plates:} Protective defenses around towers, which reduce
throughout the game.

\textbf{XP}: Experience points. These points determine your champion's
level and unlock their abilities as they progress throughout the game.

\hypertarget{sec-skill-categories}{%
\subsection{Player skills}\label{sec-skill-categories}}

\begin{enumerate}
\def\labelenumi{\arabic{enumi}.}
\item
  Mechanics

  Mechanics refer to the game-agnostic skills that players possess,
  which are fundamental to their performance across various games. These
  include:

  \begin{itemize}
  \item
    Reaction Time: The ability to quickly and effectively respond to
    game stimuli.
  \item
    Aim: Precision in targeting, crucial in shooting and action games.
  \item
    Spatial Awareness: Understanding and utilizing the in-game
    environment effectively.
  \end{itemize}
\item
  Technical Skills

  Technical skills are specific to each game and involve individualized
  abilities that vary depending on the game's mechanics and objectives.
  Key aspects include:

  \begin{itemize}
  \item
    Last-Hitting: Specific to certain strategy games, where players time
    their attacks to deal the final blow to enemy units.
  \item
    Teamfighting: Skills related to engaging in coordinated group
    battles within the game.
  \item
    Champion Depth: Expertise in using specific game characters,
    understanding their strengths, weaknesses, and optimal use cases.
  \end{itemize}

  Note: Higher mechanical ability generally enhances the speed and
  efficiency of acquiring and executing these technical skills, allowing
  for more advanced gameplay.
\item
  Strategy

  Strategy encompasses the overarching plan and approach a team takes in
  a game. It involves:

  \begin{itemize}
  \item
    Team Compositions: Deciding the combination of characters or units
    to form a balanced and effective team.
  \item
    Lane Allocations: Assigning players or resources to different areas
    of the game map.
  \item
    Teamfight Formations: Strategizing the positioning and roles of team
    members during group battles.
  \item
    Strategic Playbook: A collection of pre-planned strategies and
    approaches the team can deploy.
  \item
    Impact of Technical Skills: Teams with members possessing higher
    technical skills have access to a broader range of strategic
    options, allowing for more dynamic and adaptable gameplay.
  \end{itemize}
\item
  Tactics

  Tactics are the real-time decisions and actions taken in response to
  immediate game situations and opponent strategies. They include:

  \begin{itemize}
  \item
    Adaptability: Changing the team's approach based on the current
    state of the game.
  \item
    Opponent Analysis: Understanding and exploiting the weaknesses or
    tendencies of the opposing team.
  \item
    Tactical Execution: Effectively implementing tactics in response to
    in-game developments.
  \item
    Synergy with Technical Skills and Strategy: Teams with advanced
    technical skills and a diverse strategic playbook can deploy more
    varied and effective tactics against their opponents.
  \end{itemize}
\end{enumerate}

\hypertarget{sec-app-mm}{%
\subsection{Meta-metric tables}\label{sec-app-mm}}

\hypertarget{sec-app-disc}{%
\subsubsection{Discrimination}\label{sec-app-disc}}

Table~\ref{tbl-discrim_gold} provides discrimination scores for all gold
models while Table~\ref{tbl-discrim_dmg} provides scores for all damage
models. Note the low discrimination scores for Plus-Minus; this is a
by-product of the majority of player effects being near 0 and thus
difficult to differentiate.

\hypertarget{tbl-discrim_gold}{}
\begin{longtable}{l|lcrrrrrr}
\caption{\label{tbl-discrim_gold}Statistic discrimination values for all gold models. }\tabularnewline

\toprule
\multicolumn{1}{l}{} &  &  & \multicolumn{2}{c}{Player} & \multicolumn{2}{c}{Ally} & \multicolumn{2}{c}{Enemy} \\ 
\cmidrule(lr){4-5} \cmidrule(lr){6-7} \cmidrule(lr){8-9}
\multicolumn{1}{l}{} & Server & Time & SIDO & BA & SIDO & Plus-Minus & SIDO & Plus-Minus \\ 
\midrule\addlinespace[2.5pt]
Top & EUW & 0-7 min & 0.77 & 0.87 & 0.30 & 0.00 & 0.46 & 0.01 \\ 
 & EUW & 7-15 min & 0.82 & 0.89 & 0.33 & 0.00 & 0.55 & 0.00 \\ 
 & EUW & 15-25 min & 0.75 & 0.89 & 0.42 & 0.01 & 0.27 & 0.00 \\ 
 & KR & 0-7 min & 0.79 & 0.88 & 0.32 & 0.01 & 0.44 & 0.00 \\ 
 & KR & 7-15 min & 0.83 & 0.90 & 0.23 & 0.00 & 0.54 & 0.00 \\ 
 & KR & 15-25 min & 0.75 & 0.88 & 0.32 & 0.00 & 0.23 & 0.00 \\ 
 & NA & 0-7 min & 0.76 & 0.88 & 0.23 & 0.01 & 0.39 & 0.00 \\ 
 & NA & 7-15 min & 0.81 & 0.90 & 0.24 & 0.00 & 0.48 & 0.00 \\ 
 & NA & 15-25 min & 0.75 & 0.90 & 0.22 & 0.02 & 0.31 & 0.00 \\ 
\midrule\addlinespace[2.5pt]
Jungle & EUW & 0-7 min & 0.64 & 0.84 & 0.51 & 0.00 & 0.51 & 0.01 \\ 
 & EUW & 7-15 min & 0.74 & 0.89 & 0.50 & 0.00 & 0.59 & 0.00 \\ 
 & EUW & 15-25 min & 0.65 & 0.87 & 0.36 & 0.01 & 0.41 & 0.00 \\ 
 & KR & 0-7 min & 0.70 & 0.84 & 0.51 & 0.01 & 0.44 & 0.00 \\ 
 & KR & 7-15 min & 0.72 & 0.86 & 0.37 & 0.00 & 0.52 & 0.00 \\ 
 & KR & 15-25 min & 0.63 & 0.83 & 0.39 & 0.00 & 0.31 & 0.00 \\ 
 & NA & 0-7 min & 0.67 & 0.86 & 0.45 & 0.01 & 0.42 & 0.00 \\ 
 & NA & 7-15 min & 0.81 & 0.91 & 0.42 & 0.00 & 0.45 & 0.00 \\ 
 & NA & 15-25 min & 0.67 & 0.88 & 0.35 & 0.02 & 0.37 & 0.00 \\ 
\midrule\addlinespace[2.5pt]
Mid & EUW & 0-7 min & 0.68 & 0.82 & 0.36 & 0.00 & 0.44 & 0.01 \\ 
 & EUW & 7-15 min & 0.74 & 0.84 & 0.43 & 0.00 & 0.54 & 0.00 \\ 
 & EUW & 15-25 min & 0.70 & 0.84 & 0.34 & 0.01 & 0.27 & 0.00 \\ 
 & KR & 0-7 min & 0.76 & 0.84 & 0.19 & 0.01 & 0.46 & 0.00 \\ 
 & KR & 7-15 min & 0.80 & 0.87 & 0.31 & 0.00 & 0.51 & 0.00 \\ 
 & KR & 15-25 min & 0.74 & 0.86 & 0.37 & 0.00 & 0.32 & 0.00 \\ 
 & NA & 0-7 min & 0.75 & 0.82 & 0.25 & 0.01 & 0.34 & 0.00 \\ 
 & NA & 7-15 min & 0.76 & 0.86 & 0.26 & 0.00 & 0.40 & 0.00 \\ 
 & NA & 15-25 min & 0.71 & 0.85 & 0.29 & 0.02 & 0.33 & 0.00 \\ 
\midrule\addlinespace[2.5pt]
Bot & EUW & 0-7 min & 0.70 & 0.86 & 0.37 & 0.00 & 0.57 & 0.01 \\ 
 & EUW & 7-15 min & 0.76 & 0.87 & 0.44 & 0.00 & 0.54 & 0.00 \\ 
 & EUW & 15-25 min & 0.68 & 0.83 & 0.41 & 0.01 & 0.39 & 0.00 \\ 
 & KR & 0-7 min & 0.61 & 0.79 & 0.39 & 0.01 & 0.42 & 0.00 \\ 
 & KR & 7-15 min & 0.75 & 0.86 & 0.38 & 0.00 & 0.41 & 0.00 \\ 
 & KR & 15-25 min & 0.62 & 0.78 & 0.30 & 0.00 & 0.31 & 0.00 \\ 
 & NA & 0-7 min & 0.70 & 0.86 & 0.33 & 0.01 & 0.42 & 0.00 \\ 
 & NA & 7-15 min & 0.81 & 0.89 & 0.33 & 0.00 & 0.51 & 0.00 \\ 
 & NA & 15-25 min & 0.72 & 0.83 & 0.30 & 0.02 & 0.31 & 0.00 \\ 
\midrule\addlinespace[2.5pt]
Support & EUW & 0-7 min & 0.73 & 0.89 & 0.38 & 0.00 & 0.53 & 0.01 \\ 
 & EUW & 7-15 min & 0.78 & 0.93 & 0.31 & 0.00 & 0.54 & 0.00 \\ 
 & EUW & 15-25 min & 0.48 & 0.86 & 0.25 & 0.01 & 0.32 & 0.00 \\ 
 & KR & 0-7 min & 0.67 & 0.89 & 0.37 & 0.01 & 0.46 & 0.00 \\ 
 & KR & 7-15 min & 0.79 & 0.92 & 0.32 & 0.00 & 0.43 & 0.00 \\ 
 & KR & 15-25 min & 0.49 & 0.82 & 0.24 & 0.00 & 0.35 & 0.00 \\ 
 & NA & 0-7 min & 0.70 & 0.87 & 0.32 & 0.01 & 0.49 & 0.00 \\ 
 & NA & 7-15 min & 0.79 & 0.92 & 0.31 & 0.00 & 0.47 & 0.00 \\ 
 & NA & 15-25 min & 0.50 & 0.82 & 0.28 & 0.02 & 0.38 & 0.00 \\ 
\bottomrule
\end{longtable}

\hypertarget{tbl-discrim_dmg}{}
\begin{longtable}{l|lcrrrrrr}
\caption{\label{tbl-discrim_dmg}Statistic discrimination values for all damage models. }\tabularnewline

\toprule
\multicolumn{1}{l}{} &  &  & \multicolumn{2}{c}{Player} & \multicolumn{2}{c}{Ally} & \multicolumn{2}{c}{Enemy} \\ 
\cmidrule(lr){4-5} \cmidrule(lr){6-7} \cmidrule(lr){8-9}
\multicolumn{1}{l}{} & Server & Time & SIDO & BA & SIDO & Plus-Minus & SIDO & Plus-Minus \\ 
\midrule\addlinespace[2.5pt]
Top & EUW & 0-7 min & 0.83 & 0.93 & 0.39 & 0.00 & 0.52 & 0.00 \\ 
 & EUW & 7-15 min & 0.77 & 0.93 & 0.45 & 0.00 & 0.54 & 0.00 \\ 
 & EUW & 15-25 min & 0.64 & 0.88 & 0.46 & 0.00 & 0.41 & 0.00 \\ 
 & KR & 0-7 min & 0.82 & 0.92 & 0.29 & 0.47 & 0.57 & 0.47 \\ 
 & KR & 7-15 min & 0.82 & 0.91 & 0.31 & 0.20 & 0.51 & 0.19 \\ 
 & KR & 15-25 min & 0.64 & 0.84 & 0.28 & 0.00 & 0.30 & 0.00 \\ 
 & NA & 0-7 min & 0.83 & 0.93 & 0.31 & 0.22 & 0.50 & 0.22 \\ 
 & NA & 7-15 min & 0.80 & 0.91 & 0.29 & 0.20 & 0.53 & 0.24 \\ 
 & NA & 15-25 min & 0.72 & 0.86 & 0.36 & 0.05 & 0.50 & 0.00 \\ 
\midrule\addlinespace[2.5pt]
Jungle & EUW & 0-7 min & 0.85 & 0.90 & 0.42 & 0.00 & 0.49 & 0.00 \\ 
 & EUW & 7-15 min & 0.80 & 0.92 & 0.43 & 0.00 & 0.60 & 0.00 \\ 
 & EUW & 15-25 min & 0.73 & 0.91 & 0.48 & 0.00 & 0.46 & 0.00 \\ 
 & KR & 0-7 min & 0.82 & 0.91 & 0.42 & 0.47 & 0.38 & 0.47 \\ 
 & KR & 7-15 min & 0.77 & 0.91 & 0.34 & 0.20 & 0.49 & 0.19 \\ 
 & KR & 15-25 min & 0.68 & 0.89 & 0.19 & 0.00 & 0.32 & 0.00 \\ 
 & NA & 0-7 min & 0.81 & 0.90 & 0.34 & 0.22 & 0.48 & 0.22 \\ 
 & NA & 7-15 min & 0.81 & 0.92 & 0.39 & 0.20 & 0.54 & 0.24 \\ 
 & NA & 15-25 min & 0.73 & 0.91 & 0.37 & 0.05 & 0.44 & 0.00 \\ 
\midrule\addlinespace[2.5pt]
Mid & EUW & 0-7 min & 0.85 & 0.93 & 0.26 & 0.00 & 0.51 & 0.00 \\ 
 & EUW & 7-15 min & 0.79 & 0.90 & 0.35 & 0.00 & 0.53 & 0.00 \\ 
 & EUW & 15-25 min & 0.65 & 0.85 & 0.37 & 0.00 & 0.36 & 0.00 \\ 
 & KR & 0-7 min & 0.83 & 0.92 & 0.38 & 0.47 & 0.53 & 0.47 \\ 
 & KR & 7-15 min & 0.80 & 0.90 & 0.32 & 0.20 & 0.56 & 0.19 \\ 
 & KR & 15-25 min & 0.68 & 0.85 & 0.39 & 0.00 & 0.29 & 0.00 \\ 
 & NA & 0-7 min & 0.81 & 0.90 & 0.28 & 0.22 & 0.54 & 0.22 \\ 
 & NA & 7-15 min & 0.78 & 0.89 & 0.31 & 0.20 & 0.54 & 0.24 \\ 
 & NA & 15-25 min & 0.70 & 0.85 & 0.29 & 0.05 & 0.48 & 0.00 \\ 
\midrule\addlinespace[2.5pt]
Bot & EUW & 0-7 min & 0.78 & 0.89 & 0.39 & 0.00 & 0.49 & 0.00 \\ 
 & EUW & 7-15 min & 0.78 & 0.91 & 0.39 & 0.00 & 0.53 & 0.00 \\ 
 & EUW & 15-25 min & 0.68 & 0.85 & 0.48 & 0.00 & 0.40 & 0.00 \\ 
 & KR & 0-7 min & 0.79 & 0.89 & 0.34 & 0.47 & 0.52 & 0.47 \\ 
 & KR & 7-15 min & 0.75 & 0.88 & 0.36 & 0.20 & 0.48 & 0.19 \\ 
 & KR & 15-25 min & 0.64 & 0.80 & 0.41 & 0.00 & 0.33 & 0.00 \\ 
 & NA & 0-7 min & 0.76 & 0.90 & 0.30 & 0.22 & 0.41 & 0.22 \\ 
 & NA & 7-15 min & 0.77 & 0.91 & 0.27 & 0.20 & 0.55 & 0.24 \\ 
 & NA & 15-25 min & 0.71 & 0.86 & 0.33 & 0.05 & 0.40 & 0.00 \\ 
\midrule\addlinespace[2.5pt]
Support & EUW & 0-7 min & 0.83 & 0.96 & 0.39 & 0.00 & 0.51 & 0.00 \\ 
 & EUW & 7-15 min & 0.85 & 0.97 & 0.36 & 0.00 & 0.59 & 0.00 \\ 
 & EUW & 15-25 min & 0.80 & 0.95 & 0.31 & 0.00 & 0.38 & 0.00 \\ 
 & KR & 0-7 min & 0.82 & 0.96 & 0.35 & 0.47 & 0.49 & 0.47 \\ 
 & KR & 7-15 min & 0.87 & 0.97 & 0.36 & 0.20 & 0.54 & 0.19 \\ 
 & KR & 15-25 min & 0.80 & 0.94 & 0.23 & 0.00 & 0.42 & 0.00 \\ 
 & NA & 0-7 min & 0.83 & 0.96 & 0.42 & 0.22 & 0.46 & 0.22 \\ 
 & NA & 7-15 min & 0.87 & 0.97 & 0.41 & 0.20 & 0.55 & 0.24 \\ 
 & NA & 15-25 min & 0.84 & 0.96 & 0.28 & 0.05 & 0.49 & 0.00 \\ 
\bottomrule
\end{longtable}

\hypertarget{sec-app-ind}{%
\subsubsection{Independence}\label{sec-app-ind}}

Table~\ref{tbl-independence} provides independence scores for all
models. Table~\ref{tbl-independence-sido} provides independence scores
when considering only the SIDO models.

\hypertarget{tbl-independence}{}
\begin{longtable}{l|llllllll}
\caption{\label{tbl-independence}Statistic independence scores for all models. }\tabularnewline

\toprule
\multicolumn{1}{l}{} &  &  & \multicolumn{2}{c}{Player} & \multicolumn{2}{c}{Ally} & \multicolumn{2}{c}{Enemy} \\ 
\cmidrule(lr){4-5} \cmidrule(lr){6-7} \cmidrule(lr){8-9}
\multicolumn{1}{l}{} & statistic & Server & BA & SIDO & Plus-Minus & SIDO & Plus-Minus & SIDO \\ 
\midrule\addlinespace[2.5pt]
Top & Damage & EUW & 0.07 & 0.58 & 0.99 & 0.58 & 0.99 & 0.62 \\ 
 & Damage & KR & 0.06 & 0.60 & 0.98 & 0.51 & 1.00 & 0.54 \\ 
 & Damage & NA & 0.07 & 0.63 & 0.99 & 0.63 & 0.99 & 0.63 \\ 
 & Gold & EUW & 0.07 & 0.56 & 0.99 & 0.58 & 0.98 & 0.63 \\ 
 & Gold & KR & 0.06 & 0.60 & 0.99 & 0.51 & 0.99 & 0.54 \\ 
 & Gold & NA & 0.07 & 0.60 & 0.99 & 0.63 & 0.99 & 0.63 \\ 
\midrule\addlinespace[2.5pt]
Jungle & Damage & EUW & 0.06 & 0.65 & 0.99 & 0.66 & 0.99 & 0.56 \\ 
 & Damage & KR & 0.06 & 0.51 & 0.98 & 0.67 & 1.00 & 0.60 \\ 
 & Damage & NA & 0.06 & 0.65 & 0.99 & 0.63 & 0.99 & 0.56 \\ 
 & Gold & EUW & 0.06 & 0.64 & 1.00 & 0.58 & 0.99 & 0.55 \\ 
 & Gold & KR & 0.06 & 0.49 & 1.00 & 0.58 & 0.99 & 0.58 \\ 
 & Gold & NA & 0.06 & 0.59 & 0.98 & 0.58 & 0.99 & 0.54 \\ 
\midrule\addlinespace[2.5pt]
Mid & Damage & EUW & 0.06 & 0.57 & 0.99 & 0.68 & 1.00 & 0.61 \\ 
 & Damage & KR & 0.07 & 0.59 & 0.99 & 0.63 & 0.99 & 0.52 \\ 
 & Damage & NA & 0.07 & 0.62 & 0.99 & 0.61 & 0.99 & 0.58 \\ 
 & Gold & EUW & 0.06 & 0.54 & 0.98 & 0.66 & 0.99 & 0.61 \\ 
 & Gold & KR & 0.07 & 0.54 & 0.98 & 0.62 & 0.99 & 0.51 \\ 
 & Gold & NA & 0.07 & 0.58 & 0.99 & 0.57 & 0.99 & 0.57 \\ 
\midrule\addlinespace[2.5pt]
Bot & Damage & EUW & 0.09 & 0.56 & 0.98 & 0.64 & 0.99 & 0.57 \\ 
 & Damage & KR & 0.07 & 0.48 & 0.99 & 0.58 & 0.98 & 0.53 \\ 
 & Damage & NA & 0.07 & 0.51 & 0.98 & 0.58 & 1.00 & 0.60 \\ 
 & Gold & EUW & 0.08 & 0.53 & 0.99 & 0.65 & 0.99 & 0.57 \\ 
 & Gold & KR & 0.07 & 0.48 & 0.98 & 0.55 & 0.99 & 0.53 \\ 
 & Gold & NA & 0.07 & 0.51 & 0.99 & 0.56 & 0.98 & 0.60 \\ 
\midrule\addlinespace[2.5pt]
Support & Damage & EUW & 0.21 & 0.59 & 1.00 & 0.64 & 0.98 & 0.58 \\ 
 & Damage & KR & 0.18 & 0.52 & 0.99 & 0.64 & 0.98 & 0.50 \\ 
 & Damage & NA & 0.20 & 0.52 & 0.98 & 0.60 & 0.99 & 0.60 \\ 
 & Gold & EUW & 0.22 & 0.58 & 0.99 & 0.59 & 0.97 & 0.58 \\ 
 & Gold & KR & 0.20 & 0.58 & 0.98 & 0.56 & 0.97 & 0.49 \\ 
 & Gold & NA & 0.22 & 0.51 & 0.98 & 0.52 & 0.99 & 0.60 \\ 
\bottomrule
\end{longtable}

\hypertarget{tbl-independence-sido}{}
\begin{longtable}{l|lllllll}
\caption{\label{tbl-independence-sido}Statistic independence scores for all models. }\tabularnewline

\toprule
\multicolumn{1}{l}{} &  & \multicolumn{3}{c}{Gold} & \multicolumn{3}{c}{Damage} \\ 
\cmidrule(lr){3-5} \cmidrule(lr){6-8}
\multicolumn{1}{l}{} & Server & Player & Ally & Enemy & Player & Ally & Enemy \\ 
\midrule\addlinespace[2.5pt]
Top & EUW & 0.68 & 0.59 & 0.63 & 0.68 & 0.59 & 0.64 \\ 
 & KR & 0.70 & 0.52 & 0.55 & 0.71 & 0.52 & 0.55 \\ 
 & NA & 0.72 & 0.63 & 0.63 & 0.73 & 0.63 & 0.65 \\ 
\midrule\addlinespace[2.5pt]
Jungle & EUW & 0.77 & 0.58 & 0.55 & 0.69 & 0.66 & 0.57 \\ 
 & KR & 0.57 & 0.58 & 0.59 & 0.55 & 0.67 & 0.60 \\ 
 & NA & 0.71 & 0.59 & 0.55 & 0.69 & 0.63 & 0.57 \\ 
\midrule\addlinespace[2.5pt]
Mid & EUW & 0.69 & 0.67 & 0.61 & 0.70 & 0.68 & 0.62 \\ 
 & KR & 0.70 & 0.63 & 0.51 & 0.70 & 0.64 & 0.53 \\ 
 & NA & 0.73 & 0.58 & 0.57 & 0.71 & 0.62 & 0.59 \\ 
\midrule\addlinespace[2.5pt]
Bot & EUW & 0.63 & 0.66 & 0.58 & 0.60 & 0.64 & 0.58 \\ 
 & KR & 0.52 & 0.56 & 0.54 & 0.53 & 0.58 & 0.53 \\ 
 & NA & 0.60 & 0.57 & 0.61 & 0.59 & 0.58 & 0.61 \\ 
\midrule\addlinespace[2.5pt]
Support & EUW & 0.61 & 0.60 & 0.58 & 0.68 & 0.64 & 0.59 \\ 
 & KR & 0.64 & 0.57 & 0.49 & 0.73 & 0.64 & 0.50 \\ 
 & NA & 0.55 & 0.53 & 0.61 & 0.63 & 0.61 & 0.61 \\ 
\bottomrule
\end{longtable}

\hypertarget{sec-app-stab}{%
\subsubsection{Stability}\label{sec-app-stab}}

Table~\ref{tbl-stability-gold} provides concordance index values for all
gold models while Table~\ref{tbl-stability-dmg} provides concordance
index values for all damage models.

\hypertarget{tbl-stability-gold}{}
\begin{longtable}{l|lcrrrr}
\caption{\label{tbl-stability-gold}Statistic stability scores for SIDO and BA gold models. }\tabularnewline

\toprule
\multicolumn{1}{l}{} &  &  & \multicolumn{2}{c}{Player} & Ally & Enemy \\ 
\cmidrule(lr){4-5} \cmidrule(lr){6-6} \cmidrule(lr){7-7}
\multicolumn{1}{l}{} & Server & Time & SIDO & BA & SIDO & SIDO \\ 
\midrule\addlinespace[2.5pt]
Top & EUW & 0-7 min & 0.73 & 0.76 & 0.54 & 0.58 \\ 
 & EUW & 7-15 min & 0.72 & 0.77 & 0.55 & 0.60 \\ 
 & EUW & 15-25 min & 0.68 & 0.80 & 0.55 & 0.55 \\ 
 & KR & 0-7 min & 0.75 & 0.77 & 0.50 & 0.59 \\ 
 & KR & 7-15 min & 0.72 & 0.76 & 0.48 & 0.65 \\ 
 & KR & 15-25 min & 0.72 & 0.78 & 0.50 & 0.53 \\ 
 & NA & 0-7 min & 0.65 & 0.77 & 0.54 & 0.59 \\ 
 & NA & 7-15 min & 0.73 & 0.82 & 0.56 & 0.64 \\ 
 & NA & 15-25 min & 0.68 & 0.80 & 0.56 & 0.62 \\ 
\midrule\addlinespace[2.5pt]
Jungle & EUW & 0-7 min & 0.66 & 0.73 & 0.55 & 0.60 \\ 
 & EUW & 7-15 min & 0.70 & 0.77 & 0.65 & 0.59 \\ 
 & EUW & 15-25 min & 0.68 & 0.73 & 0.56 & 0.57 \\ 
 & KR & 0-7 min & 0.62 & 0.62 & 0.59 & 0.59 \\ 
 & KR & 7-15 min & 0.66 & 0.69 & 0.53 & 0.62 \\ 
 & KR & 15-25 min & 0.63 & 0.67 & 0.51 & 0.50 \\ 
 & NA & 0-7 min & 0.59 & 0.68 & 0.58 & 0.59 \\ 
 & NA & 7-15 min & 0.70 & 0.76 & 0.55 & 0.57 \\ 
 & NA & 15-25 min & 0.67 & 0.75 & 0.54 & 0.61 \\ 
\midrule\addlinespace[2.5pt]
Mid & EUW & 0-7 min & 0.64 & 0.70 & 0.55 & 0.57 \\ 
 & EUW & 7-15 min & 0.68 & 0.71 & 0.47 & 0.61 \\ 
 & EUW & 15-25 min & 0.66 & 0.70 & 0.55 & 0.53 \\ 
 & KR & 0-7 min & 0.71 & 0.71 & 0.57 & 0.62 \\ 
 & KR & 7-15 min & 0.66 & 0.73 & 0.55 & 0.60 \\ 
 & KR & 15-25 min & 0.68 & 0.75 & 0.56 & 0.53 \\ 
 & NA & 0-7 min & 0.70 & 0.76 & 0.55 & 0.63 \\ 
 & NA & 7-15 min & 0.68 & 0.74 & 0.54 & 0.60 \\ 
 & NA & 15-25 min & 0.67 & 0.74 & 0.50 & 0.55 \\ 
\midrule\addlinespace[2.5pt]
Bot & EUW & 0-7 min & 0.67 & 0.75 & 0.53 & 0.61 \\ 
 & EUW & 7-15 min & 0.73 & 0.75 & 0.60 & 0.62 \\ 
 & EUW & 15-25 min & 0.72 & 0.73 & 0.59 & 0.50 \\ 
 & KR & 0-7 min & 0.64 & 0.65 & 0.62 & 0.61 \\ 
 & KR & 7-15 min & 0.69 & 0.71 & 0.60 & 0.56 \\ 
 & KR & 15-25 min & 0.73 & 0.71 & 0.56 & 0.53 \\ 
 & NA & 0-7 min & 0.66 & 0.71 & 0.53 & 0.57 \\ 
 & NA & 7-15 min & 0.70 & 0.76 & 0.59 & 0.54 \\ 
 & NA & 15-25 min & 0.71 & 0.74 & 0.57 & 0.58 \\ 
\midrule\addlinespace[2.5pt]
Support & EUW & 0-7 min & 0.66 & 0.74 & 0.55 & 0.63 \\ 
 & EUW & 7-15 min & 0.70 & 0.78 & 0.54 & 0.57 \\ 
 & EUW & 15-25 min & 0.60 & 0.71 & 0.56 & 0.54 \\ 
 & KR & 0-7 min & 0.62 & 0.68 & 0.50 & 0.58 \\ 
 & KR & 7-15 min & 0.71 & 0.74 & 0.57 & 0.56 \\ 
 & KR & 15-25 min & 0.53 & 0.64 & 0.52 & 0.53 \\ 
 & NA & 0-7 min & 0.67 & 0.72 & 0.54 & 0.62 \\ 
 & NA & 7-15 min & 0.70 & 0.72 & 0.60 & 0.59 \\ 
 & NA & 15-25 min & 0.56 & 0.65 & 0.49 & 0.53 \\ 
\bottomrule
\end{longtable}

\hypertarget{tbl-stability-dmg}{}
\begin{longtable}{l|lcrrrr}
\caption{\label{tbl-stability-dmg}Statistic stability scores for SIDO and BA damage models. }\tabularnewline

\toprule
\multicolumn{1}{l}{} &  &  & \multicolumn{2}{c}{Player} & Ally & Enemy \\ 
\cmidrule(lr){4-5} \cmidrule(lr){6-6} \cmidrule(lr){7-7}
\multicolumn{1}{l}{} & Server & Time & SIDO & BA & SIDO & SIDO \\ 
\midrule\addlinespace[2.5pt]
Top & EUW & 0-7 min & 0.75 & 0.81 & 0.58 & 0.57 \\ 
 & EUW & 7-15 min & 0.69 & 0.79 & 0.56 & 0.60 \\ 
 & EUW & 15-25 min & 0.63 & 0.73 & 0.60 & 0.57 \\ 
 & KR & 0-7 min & 0.75 & 0.78 & 0.54 & 0.63 \\ 
 & KR & 7-15 min & 0.71 & 0.79 & 0.55 & 0.63 \\ 
 & KR & 15-25 min & 0.66 & 0.75 & 0.53 & 0.60 \\ 
 & NA & 0-7 min & 0.68 & 0.78 & 0.55 & 0.64 \\ 
 & NA & 7-15 min & 0.67 & 0.74 & 0.57 & 0.56 \\ 
 & NA & 15-25 min & 0.64 & 0.78 & 0.55 & 0.59 \\ 
\midrule\addlinespace[2.5pt]
Jungle & EUW & 0-7 min & 0.72 & 0.76 & 0.58 & 0.59 \\ 
 & EUW & 7-15 min & 0.70 & 0.77 & 0.60 & 0.58 \\ 
 & EUW & 15-25 min & 0.68 & 0.74 & 0.58 & 0.58 \\ 
 & KR & 0-7 min & 0.67 & 0.69 & 0.55 & 0.56 \\ 
 & KR & 7-15 min & 0.67 & 0.70 & 0.55 & 0.62 \\ 
 & KR & 15-25 min & 0.66 & 0.71 & 0.50 & 0.53 \\ 
 & NA & 0-7 min & 0.70 & 0.73 & 0.55 & 0.59 \\ 
 & NA & 7-15 min & 0.70 & 0.73 & 0.51 & 0.61 \\ 
 & NA & 15-25 min & 0.64 & 0.71 & 0.58 & 0.50 \\ 
\midrule\addlinespace[2.5pt]
Mid & EUW & 0-7 min & 0.76 & 0.78 & 0.60 & 0.56 \\ 
 & EUW & 7-15 min & 0.71 & 0.74 & 0.50 & 0.60 \\ 
 & EUW & 15-25 min & 0.65 & 0.71 & 0.54 & 0.57 \\ 
 & KR & 0-7 min & 0.74 & 0.78 & 0.60 & 0.55 \\ 
 & KR & 7-15 min & 0.71 & 0.75 & 0.55 & 0.63 \\ 
 & KR & 15-25 min & 0.65 & 0.71 & 0.53 & 0.50 \\ 
 & NA & 0-7 min & 0.71 & 0.77 & 0.58 & 0.59 \\ 
 & NA & 7-15 min & 0.62 & 0.74 & 0.55 & 0.63 \\ 
 & NA & 15-25 min & 0.64 & 0.69 & 0.54 & 0.57 \\ 
\midrule\addlinespace[2.5pt]
Bot & EUW & 0-7 min & 0.69 & 0.73 & 0.59 & 0.62 \\ 
 & EUW & 7-15 min & 0.69 & 0.78 & 0.58 & 0.63 \\ 
 & EUW & 15-25 min & 0.63 & 0.72 & 0.60 & 0.54 \\ 
 & KR & 0-7 min & 0.72 & 0.75 & 0.55 & 0.65 \\ 
 & KR & 7-15 min & 0.69 & 0.75 & 0.57 & 0.61 \\ 
 & KR & 15-25 min & 0.64 & 0.71 & 0.55 & 0.54 \\ 
 & NA & 0-7 min & 0.63 & 0.72 & 0.51 & 0.61 \\ 
 & NA & 7-15 min & 0.64 & 0.75 & 0.61 & 0.57 \\ 
 & NA & 15-25 min & 0.64 & 0.72 & 0.57 & 0.59 \\ 
\midrule\addlinespace[2.5pt]
Support & EUW & 0-7 min & 0.71 & 0.78 & 0.54 & 0.60 \\ 
 & EUW & 7-15 min & 0.68 & 0.76 & 0.55 & 0.59 \\ 
 & EUW & 15-25 min & 0.67 & 0.78 & 0.50 & 0.60 \\ 
 & KR & 0-7 min & 0.70 & 0.74 & 0.54 & 0.63 \\ 
 & KR & 7-15 min & 0.71 & 0.77 & 0.57 & 0.65 \\ 
 & KR & 15-25 min & 0.66 & 0.74 & 0.55 & 0.61 \\ 
 & NA & 0-7 min & 0.71 & 0.72 & 0.50 & 0.62 \\ 
 & NA & 7-15 min & 0.72 & 0.70 & 0.56 & 0.64 \\ 
 & NA & 15-25 min & 0.60 & 0.70 & 0.50 & 0.54 \\ 
\bottomrule
\end{longtable}

\end{document}